\documentclass[reqno]{amsart}

\usepackage{amsmath,amsfonts}
\usepackage{epsfig}
\usepackage{graphicx}

\def\alphaset{{\mathfrak A}}

\def\CSob{C_{Sob}}

\def\fjkl{f_{j_1\dots j_q}}
\def\Gspace{{\mathfrak G}}

\def\opDelta{\widehat{\Delta}}
\def\opB{\widehat{B}}

\def\e{\epsilon}

\def\tr{{\rm Tr}}

\def\bra{\big\langle}
\def\ket{\big\rangle}

\def\C{{\mathbb C}}
\def\N{{\mathbb N}}

\def\R{{\mathbb R}}
\def\Z{{\mathbb Z}}

\def\uu{{\underline{u}}}

\def\ux{{\underline{x}}}
\def\uxi{{\underline{\xi}}}

\def\uy{{\underline{y}}}

\def\frh{{\mathfrak h}}
\def\frH{{\mathfrak H}}

\def\cH{{\mathcal H}}
\def\cI{{\mathcal I}}
\def\cK{{\mathcal K}}

\def\1{{\bf 1}}

\def\eqnn{\begin{eqnarray*}}
\def\eeqnn{\end{eqnarray*}}
\def\eqn{\begin{eqnarray}}
\def\eeqn{\end{eqnarray}}

\def\prf{\begin{proof}}
\def\endprf{\end{proof}}

\theoremstyle{plain}
\newtheorem{theorem}{Theorem}[section]

\newtheorem{remark}[theorem]{Remark}

\numberwithin{equation}{section}

\begin{document}

\parskip=8pt

\title[Global well-posedness for the GP hierarchy]
{Higher order energy conservation and global wellposedness of solutions for
Gross-Pitaevskii hierarchies}

\author[T. Chen]{Thomas Chen}
\address{T. Chen,  
Department of Mathematics, University of Texas at Austin.}
\email{tc@math.utexas.edu}

\author[N. Pavlovi\'{c}]{Nata\v{s}a Pavlovi\'{c}}
\address{N. Pavlovi\'{c},  
Department of Mathematics, University of Texas at Austin.}
\email{natasa@math.utexas.edu}

%\date{\mscrptdate}

\begin{abstract}
We consider the cubic and quintic Gross-Pitaevskii (GP) hierarchies in $d$ dimensions, 
for focusing and defocusing interactions. 
We introduce new higher order energy functionals and prove that they
are conserved for solutions of energy subcritical defocusing, and
$L^2$ subcritical (de)focusing GP hierarchies, in spaces also 
used by Erd\"os, Schlein and Yau in  \cite{esy1,esy2}.
By use of this tool, we prove a priori $H^1$ bounds for positive semidefinite solutions
in those spaces. 
Moreover, we obtain global well-posedness results for positive semidefinite solutions 
in the spaces studied in the works of Klainerman and Machedon, \cite{klma}, and in \cite{chpa2}.
As part of our analysis, we prove generalizations of Sobolev and Gagliardo-Nirenberg inequalities
for density matrices.
\end{abstract}

\maketitle

\section{Introduction}

In recent years, there has been impressive progress 
related to the derivation of nonlinear dispersive PDEs,
such as the nonlinear Schr\"odinger (NLS) or nonlinear Hartree (NLH) equations, 
from many body quantum dynamics,
%as effective theories describing 
%the mean field dynamics of weakly interacting Bose gases, 
see  \cite{esy1,esy2,ey,kiscst,klma,rosc} and the references therein,
and also \cite{adgote,anasig,eesy,frgrsc,frknpi,frknsc,grma,grmama,he,sp}.
Closely related to this research field is the mathematical study of Bose-Einstein condensation 
in systems of interacting bosons, where we refer to the highly influencial works
\cite{ailisesoyn,lise,lisesoyn,liseyn} and the references therein.

\subsection{The Gross-Pitaevkii limit for Bose gases}  
\label{subsec-exsol-GP-0}
In the landmark works \cite{esy1,esy2,ey}, Erd\"os, Schlein, and Yau 
developed a powerful method to derive the NLS as a dynamical Gross-Pitaevskii limit 
of an interacting Bose gas. 
For the convenience of the reader, and as a preparation for our discussion
below, we will here outline some of the 
main steps,
following \cite{esy1,esy2,ey}.  For 
the derivation of the defocusing quintic NLS from a system of
bosons with repelling three body interactions, we refer to \cite{chpa}.   

\subsubsection{From $N$-body Schr\"odinger to BBGKY hierarchy}
We consider a system of $N$ bosons in $\R^d$ where 
$\psi_{N} \in L^2(\R^{dN})$ denotes its wave function. 
To comply with Bose-Einstein statistics,  
$\psi_{N}$ is invariant under the permutation of particle variables,  
\begin{equation}\label{sym}
	\psi_N(x_{\pi( 1)},x_{\pi (2)},...,x_{\pi (N)}) \, = \, \psi_N(x_1, x_2,..., x_N) \
	\; \; \; \; \; \; \; \; \forall \pi \in S_{N} \,,
\end{equation} 
where  $S_{N}$ is the $N$-th symmetric group.
We denote by $L_{s}^{2}(\mathbb R^{dN})$ the subspace  of 
$L^2(\R^{dN})$ of elements obeying \eqref{sym}.
%$L_{s}^{2}(\mathbb R^{dN}):=\{\psi_N\in L^{2}(\mathbb R^{dN})| \, \psi_N {\rm \; satisfies \;}\eqref{sym}\}$. 
The dynamics of the system is determined by the $N$-body Schr\"odinger equation
\begin{equation}\label{ham1}
i\partial_{t}\psi_{N} \, = \, H_{N}\psi_{N} \,.
\end{equation} 
The Hamiltonian $H_{N}$ is assumed to be a self-adjoint operator acting on the 
Hilbert space $L_{s}^2(\mathbb R^{dN})$, given by 
\begin{equation}\label{ham2}
	H_{N} \, = \, \sum_{j=1}^{N}(-\Delta_{x_{j}})+\frac1N\sum_{1\leq i<j \leq N}V_N(x_{i}-x_{j}),
\end{equation}
where $V_N(x)=N^{d\beta}V(N^\beta x)$ with 
$V\in W^{r,s}(\R^d)$ spherically symmetric, for some suitable $r$, $s$, 
and for $\beta\in(0,1]$ sufficiently small\footnote{When $\beta = 1$, the Hamiltonian \eqref{ham2} is 
called the Gross-Pitaevskii Hamiltonian. Due to the factor
$\frac{1}{N}$ in front of the interaction potential, \eqref{ham2} can be formally interpreted as 
a mean field Hamiltonian. However, it should be noted that \eqref{ham2}  in fact
describes a very dilute gas, where
interactions among particles are very rare and strong, while in a mean field
Hamiltonian each particle usually reacts with all other particles via a very weak potential. 
However, one can still apply to \eqref{ham2} similar mathematical methods 
as in the case of a mean field potential.}. 

Since the Schr\"{o}dinger equation \eqref{ham1} is linear and $H_N$ self-adjoint,  the global 
well-posedness of solutions is satisfied. 
However, due to the exceedingly large number of degrees of freedom of order $O(N)$
(varying from $N\sim10^3$ for samples of 
very dilute Bose-Einstein gases, to $N\sim10^{30}$ in boson stars),
it is very difficult to understand qualitative and quantitative properties of the 
bulk dynamics of the system, if directly based on the solution of \eqref{ham1}.
%one is usually interested in qualitative and quantitative properties of the solution, 
%which are hard to extract in physically relevant cases when 
%number of particles $N$ is very large (e.g. it varies from $10^3$ for very dilute
%Bose-Einstein samples, to $10^{30}$ in stars). 
It is often much more informative to study coarse grained, effective properties of the system
obtained from averaging over large $N$. 
%Further simplifications are related to obtaining a macroscopic behavior in the limit
In addition, taking the limit $N \rightarrow \infty$ for an appropriate scaling
typically leads to macroscopic
effective theories, or mean-field theories 
%(the nonlinear Schr\"odinger equation in the mean-field limit considered here), 
which are expected to accurately describe properties of the underlying physical
system with very large, but finite $N$. 

To perform the infinite particle number limit $N \rightarrow \infty$,
we consider the strategy developed in \cite{esy1,esy2}, which can be described as follows. 
First of all, one introduces the density matrix
$$\gamma_{N}(t, \ux_N, \ux_N')=\psi_{N}(t, \ux_N)\overline{\psi_{N}(t,\ux_N')}$$
where $\ux_N=(x_1,x_2,..., x_N)$ and $\ux_N'=(x_{1}^{\prime}, x_{2}^{\prime},..., x_{N}^{\prime})$.
Furthermore, one considers the associated sequence of 
$k$-particle marginal density matrices\footnote{The $k$-particle marginal density matrices 
play a key role in the analysis of the system as $N \rightarrow \infty$, because for every fixed $k$, 
$\gamma^{(k)}_N$ can have a well defined limit.} $\gamma_{N}^{(k)}(t)$, for $k=1,\dots,N$, 
as the partial traces of $\gamma_{N}$ 
over the degrees of freedom associated to the last $(N-k)$ particle variables,
$$\gamma_{N}^{(k)}=\tr_{k+1, k+2,...,N}|\psi_{N}\rangle\langle\psi_{N}| \,.$$
Here, $\tr_{k+1, k+2,...,N}$ denotes the partial trace with respect to the particles indexed
by $k+1, k+2,..., N$. 
Accordingly, $\gamma_{N}^{(k)}$ is explicitly given by 
%as the non-negative trace class operator 
%on $L_{s}^{2}(\mathbb R^{dk})$ with kernel given by
\eqn
	\gamma_{N}^{(k)}(\ux_k,\ux_k')
	&=& \int d\ux_{N-k} \gamma_{N}(\ux_k, \ux_{N-k};\ux_k', \ux_{N-k})
	\nonumber\\
	\label{reduced}
	&=&\int d\ux_{N-k}\overline{ \psi_{N}(\ux_k, \ux_{N-k}) }  \psi_{N}(\ux_k', \ux_{N-k}) \,.
\eeqn
It is clear from the definitions that the property of {\em admissibility} holds,
\eqn
	\gamma^{(k)}_N \, = \, \tr_{k+1}\gamma^{(k+1)}_N 
	\; \; \; \; \;  , \; \; \; \; k\, = \, 1,\dots,N-1 \,,
\eeqn 
and that 
$\tr \gamma_{N}^{(k)}=\|\psi_N\|_{L_{s}^{2}(\mathbb R^{dN})}^2=1$ for all $N$, 
and all $k=1, 2, ..., N$. 

Moreover, $\gamma_N^{(k)}\geq0$ is positive semidefinite as an operator 
${\mathcal S}(\R^{kd})\times{\mathcal S}(\R^{kd})\rightarrow\C$, 
$(f,g)\mapsto \int d\ux d\ux'f(\ux)\gamma(\ux;\ux')\overline{g(\ux')}$.

The time evolution of the density matrix $\gamma_{N}$ is determined by the Heisenberg equation
\begin{equation}\label{von}
	i\partial_{t}\gamma_{N}(t) \, = \, [H_{N}, \gamma_N(t)] \, ,
\end{equation}
which has the explicit form
\eqn
	i\partial_{t}\gamma_{N}(t,\ux_N,\ux_N')
	&=&-(\Delta_{\ux_N}-\Delta_{ \ux_N' })\gamma_{N}(t,\ux_N, \ux_N') 
	\label{von2}\\
	&&+\frac{1}{N}\sum_{1\leq i<j \leq N}[V_N(x_i-x_j)-V_N(x_i^{\prime}-x_{j}^{\prime})]
	\gamma_{N}(t, \ux_N,\ux_N') \,.
	\nonumber
\eeqn
%expressed in terms of the associated integral kernel.
Accordingly, the $k$-particle marginals satisfy the BBGKY hierarchy 
\eqn\label{BBGKY}
	\lefteqn{
	i\partial_{t}\gamma^{(k)}(t,\ux_k;\ux_k')
	 \, = \, 
	-(\Delta_{\ux_k}-\Delta_{ \ux_k'})\gamma^{(k)}(t,\ux_k,\ux_k')
	}
	\nonumber\\
	&&
	+ \frac{1}{N}\sum_{1\leq i<j \leq k}[V_N(x_i-x_j)-V_N(x_i^{\prime}-x_{j}')]
	\gamma^{(k)}(t, \ux_{k};\ux_{k}') 
	\label{eq-bbgky-1}\\
	&&+\frac{N-k}{N}\sum_{i=1}^{k}\int dx_{k+1}[V_N(x_i-x_{k+1})-V_N(x_i^{\prime}-x_{k+1})]
	\label{eq-bbgky-2}\\
	&&\quad\quad\quad\quad\quad\quad\quad\quad\quad\quad\quad\quad\quad\quad\quad\quad
	\gamma^{(k+1)}(t, \ux_{k},x_{k+1};\ux_{k},x_{k+1}')
	\nonumber
\eeqn
	%\\
	%&&+\frac{(N-k)(N-k-1)}{N} \int dx_{k+1}dx_{k+2}[V_N(x_{k+1}-x_{k+2})-V_N(x_{k+1}'-x_{k+2})]
	%\nonumber\\
	%&&\quad\quad\quad\quad\quad\quad\quad\quad\quad\quad\quad\quad\quad\quad\quad\quad
	%\gamma^{(k+2)}(t, \ux_{k+1},x_{k+2};\ux_{k+1}',x_{k+2}) \,
	%\label{eq-bbgky-3}
where $\Delta_{\ux_k}:=\sum_{j=1}^{k}\Delta_{x_j}$, and similarly for $\Delta_{\ux_k'}$. 
We note that the number of terms in (\ref{eq-bbgky-1}) is $\approx \frac{k^2}{N}\rightarrow0$, 
and the number of terms in (\ref{eq-bbgky-2}) is $\frac{k(N-k)}{N}\rightarrow k$ 
as $N\rightarrow \infty$. Accordingly, for fixed $k$, (\ref{eq-bbgky-1}) disappears in the limit 
$N\rightarrow\infty$ described below, while (\ref{eq-bbgky-2}) survives.
 
\subsubsection{From BBGKY hierarchy to GP hierarchy.}
A crucial step in this analysis is the limit $N\rightarrow\infty$,
and the extraction of closed equations for the time evolution of 
${\mbox{lim}}_{N \rightarrow \infty} \gamma^{(k)}_N$. 
It is proven in \cite{esy1,esy2,ey} that, for asymptotically factorizing initial data,
and a suitable weak topology on the space of marginal
density matrices, one can extract convergent subsequences
$\gamma^{(k)}_N\rightarrow\gamma^{(k)}$ as $N\rightarrow\infty$, for   $k\in\N$, which satisfy the 
%As shown in \cite{esy1,esy2,ey}, 
%it is possible to extract convergent subsequences
%$\gamma^{(k)}_N\rightarrow\gamma^{(k)}$, for $k\in\N$ and  $N\rightarrow\infty$,
%in a suitable topology on the space of marginal
%density matrices. The limiting sequence of density matrices satisfies
the infinite limiting hierarchy  
\eqn\label{eq-GP-0-0}
	i\partial_{t}\gamma^{(k)}(t,\ux_k;\ux_k')
	&=&
 	- \, (\Delta_{\ux_k}-\Delta_{\ux_k'})\gamma^{(k)}(t,\ux_k;\ux_k')
	\label{GP}\\
	&&+ \, b_0 \sum_{j=1}^{k}   \left(B_{j, k+1} 
	\gamma^{k+1}\right)(t,\ux_k  ; \ux_k' )  \,, 
	\nonumber
\eeqn
which is referred to as the {\em Gross-Pitaevskii (GP) hierarchy}.
Here,
\eqn
	\lefteqn{
	(B_{j, k+1} \gamma^{k+1})(t,  \ux_k ; \ux_k' )
	}
	\nonumber\\
	&:=&
	\int dx_{k+1}dx_{k+1}'[\delta(x_j-x_{k+1})\delta(x_{j}-x_{k+1}^{\prime})-\delta(x_j^{\prime}-x_{k+1})
	\delta(x_{j}^{\prime}-x_{k+1}^{\prime})]
	\nonumber\\
	&&\quad\quad\quad\quad\quad\quad\quad\quad\quad\quad\quad\quad\quad\quad\quad\quad
	\gamma^{(k+1)}(t,\ux_k, x_{k+1};\ux_k', x'_{k+1}) \,.
	\nonumber 
\eeqn
The coefficient  $b_0$ is the {\em scattering length} if $\beta=1$ (see \cite{esy1,lisesoyn} for the definition),
and $b_0=\int V(x) dx$ if $\beta<1$ (corresponding to the Born approximation of the scattering length).
For $\beta<1$, the interaction term is obtained from the weak limit $V_N(x)\rightarrow b_0\delta(x)$ 
in  (\ref{eq-bbgky-2}) as 
$N\rightarrow\infty$. The proof for the case $\beta=1$ is much more difficult, and the derivation
of the scattering length in this context is a breakthrough result obtained in \cite{esy1,esy2}.
For notational convenience, we will mostly set $b_0=1$ in the sequel.

Some key properties satisfied by the solutions of the GP hierarchy are:
\begin{itemize}
\item
The solution of the GP hierarchy obtained in \cite{esy1,esy2} exists {\em globally} in $t$.
\item
It satisfies the property of admissibility,
\eqn
	\gamma^{(k)} \, = \, \tr_{k+1}\gamma^{(k+1)} 
	\; \; \; \; , \; \; \; \; \forall \; k\in\N \,,
\eeqn 
which is inherited from the system at finite $N$.
\item
There exists a constant $C_0$ depending on the initial data only, such that
the {\em a priori energy bound}
\eqn\label{eq-ESY-apriori-enbd-0}
	\tr( \, | S^{(k,1)}\gamma^{(k)}(t)| \, ) \, < \, C_0^k
\eeqn
is satisfied for all $k\in \N$, and for all $t\in\R$, where
\eqn
	S^{(k,\alpha)} \, := \, 
	\prod_{j=1}^k \langle\nabla_{x_j}\rangle^\alpha\langle\nabla_{x_j'}\rangle^\alpha \,.
\eeqn
This is obtained from energy conservation in the original $N$-body Schr\"{o}dinger system.
\item
Solutions of the GP hierarchy are studied in   spaces
of $k$-particle marginals $\{\gamma^{(k)} \, | \, \|\gamma^{(k)}\|_{\frh^1}  \, < \, \infty\}$ with norms 
\eqn\label{eq-frhnorm-def-0}
	\|\gamma^{(k)}\|_{\frh^{\alpha}} \, := \,  \tr (|S^{(k,\alpha)}\gamma^{(k)}|)   \,.
\eeqn
This is in agreement with the a priori bounds \eqref{eq-ESY-apriori-enbd-0}.
\end{itemize}

\subsubsection{Factorized solutions of GP and NLS}

The NLS emerges as the mean field dynamics of the Bose gas for the very special
subclass of solutions of the GP hierarchy that are {\em factorized}.  
Factorized  $k-$particle marginals at time $t=0$ have the form
$$\gamma_0^{(k)}(\ux_k;\ux_k')=\prod_{j=1}^{k}\phi_0( x_{j})\overline{\phi_0( x_{j}^{\prime}})\,,$$
where we assume that $\phi_0\in H^1(\R^d)$. One can easily verify that the 
solution of the GP hierarchy remains factorized for all $t\in I\subseteq\R$,
$$\gamma^{(k)}(t,\ux_k;\ux_k')=\prod_{j=1}^{k}\phi(t,x_{j})\overline{\phi(t,x_{j}^{\prime})} \,,$$
and that for the GP hierarchy given in \eqref{eq-GP-0-0} with
$b_0=1$,  $\phi(t)\in H^1(\R^d)$ solves the defocusing cubic NLS,
\eqn
	i\partial_t\phi \, = \, - \Delta_x \phi \, + \, |\phi|^2\phi\,,
\eeqn 
for $t\in I\subseteq\R$, and $\phi(0)=\phi_0\in H^1(\R^d)$.

\subsubsection{Uniqueness of solutions of GP hierarchies.}

While the existence of factorized solutions can be easily verified in the manner
outlined above, the  proof of the {\em uniqueness of solutions} of the GP hierarchy
(which encompass non-factorized solutions) is the most difficult part in 
this analysis. The proof of uniqueness of solutions to the GP hierarchy 
was originally achieved by Erd\"os, Schlein and Yau in \cite{esy1,esy2,ey} in the space 
 $\{\gamma^{(k)} \, | \, \|\gamma^{(k)}\|_{\frh^1}  \, < \, \infty\}$,
for which the authors developed highly sophisticated Feynman graph expansion methods. 
 
In \cite{klma}, Klainerman and Machedon introduced an alternative method for proving uniqueness  
in a  space of density matrices defined by the Hilbert-Schmidt type Sobolev norms 
\eqn\label{eq-gamma-norm-def-0-1}
	\| \gamma^{(k)} \|_{H^\alpha_k} \, := \,  ( \, \tr( \, | S^{(k,\alpha)} \gamma^{(k)} |^2 \, ) \, )^{\frac12}  
	\, < \, \infty \,.
\eeqn 
While this is a different (strictly larger) space of marginal density matrices than
the one considered by Erd\"os, Schlein, and Yau, \cite{esy1,esy2},
the authors of \cite{klma} impose an additional a priori condition on  
space-time norms of the form
\eqn \label{intro-KMbound} 
     \|B_{j;k+1} \gamma^{(k+1)}\|_{L^2_t H^1_k}  \, < \, C^k \, , 
\eeqn
for some arbitrary but finite $C$ independent of $k$. 
The strategy in \cite{klma} developed to prove the uniqueness of solutions of the GP hierarchy
\eqref{eq-GP-0-0}  in $d=3$ involves the use of certain space-time bounds on density matrices 
(of generalized Strichartz type), 
and crucially employs the reformulation of a combinatorial result
in \cite{esy1,esy2} into a ``board game'' argument.
The  latter is used to organize the Duhamel expansion of solutions of the GP hierarchy
into equivalence classes of terms which leads to a significant reduction of 
the complexity of the problem. 

Subsequently, Kirkpatrick, Schlein, and Staffilani proved in \cite{kiscst} that the a priori 
spacetime bound \eqref{intro-KMbound}  
is satisfied for the cubic GP hierarchy in $d=2$, locally in time.
Their argument is based on the conservation of
energy in the original $N$-body Schr\"odinger system,
and a related a priori $H^1$-bounds for the BBGKY hierarchy in the limit $N\rightarrow\infty$
derived in \cite{esy1,esy2}, 
combined with a generalized Sobolev inequality for density matrices.

\subsection{On the Cauchy problem for GP hierarchies}

It is currently not known how to rigorously derive a GP hierarchy from the
$N\rightarrow\infty$ limit of a BBGKY hierarchy with $L^2$-supercritical, attractive
interactions.
Nevertheless, 
we have begun in \cite{chpa2} to   adopt  as our starting point the level of GP hierarchies,
and to study the well-posedness of the Cauchy problem for 
systems with both focusing and defocusing interactions.
Accordingly, the corresponding GP hierarchies are referred to as   {\em cubic},
{\em quintic}, {\em focusing}, or {\em defocusing GP hierarchies}, 
depending on the type of the NLS governing the solutions obtained from
factorized initial conditions. 

In \cite{chpa2}, we introduced the following topology on the Banach space of sequences
of $k$-particle marginal density matrices 
\eqn\label{bigG}
	\Gspace \, = \, \{ \, \Gamma \, = \, ( \, \gamma^{(k)}(x_1,\dots,x_k;x_1',\dots,x_k') \, )_{k\in\N} 
	\, | \,
	\tr \gamma^{(k)} \, < \, \infty \, \} \,.
\eeqn
Given $\xi>0$, we defined the space
\eqn
	\cH_\xi^\alpha \, = \, \{\Gamma \, | \, \| \, \Gamma \, \|_{\cH_\xi^\alpha} \, < \, \infty \, \}
\eeqn
with the norm
\eqn\label{eq-KM-aprioriassumpt-0-1}
	\| \, \Gamma \, \|_{\cH_\xi^\alpha} \, := \,
	\sum_{k\in\N} \xi^k \, \| \, \gamma^{(k)} \, \|_{H^\alpha} \,,
\eeqn
where
\eqn\label{eq-gamma-norm-def-0-0}
	\| \gamma^{(k)} \|_{H^\alpha_k} & := &  \tr( \, | S^{(k,\alpha)} \gamma^{(k)} |^2 \, )^{\frac12}  
\eeqn  
is the norm \eqref{eq-gamma-norm-def-0-1} considered in \cite{klma}.  
If $\Gamma\in\cH_\xi^\alpha$, then $\xi^{-1}$ an upper bound on the typical $H^\alpha$-energy per particle;
this notion is made precise in \cite{chpa2}.
We note that small energy results are characterized by large $\xi>1$, while results valid without any upper
bound on the size of the energy can be proven for arbitrarily small values of $\xi>0$; in the latter case, one can
assume $0<\xi<1$ without any loss of generality.
The parameter $\alpha$ determines the regularity of the solution.

In \cite{chpa2},  we prove the local well-posedness of solutions for 
energy subcritical focusing and defocusing  cubic 
and quintic GP hierarchies in a subspace of $\cH_\xi^\alpha$  
defined by a condition related to \eqref{intro-KMbound}\footnote{
The parameter $\alpha$ determines the regularity of the solution (e.g. 
for cubic GP when $d=3$, $\alpha \in [1, \infty)$).}.  
The precise formulation is given in Theorem \ref{thm-localwp-TTstar-1-2} below.
Our result 
is obtained from a Picard fixed point argument, and holds for various dimensions $d$,
without any requirement on factorization.
%More precisely we prove existence of solutions (not necessarily factorized)  
%which satisfy a space-time bound \eqref{intro-KMbound}
%that was assumed in the work of Klainerman and Machedon \cite{klma}.
%which is expressed in our notation by the condition\footnote{In fact, the inequality \eqref{uniq-spacetimebd} 
%corresponds to an inequality of Strichartz type.}
%\eqn \label{uniq-spacetimebd} 
%\|\opB\Gamma\|_{L^1_{t\in I}\cH^{1}_{\xi}}<\infty,
%\eeqn
%for some $\xi>0$. 
The parameter $\xi>0$ is determined by the initial condition, and it sets the energy scale of 
the given Cauchy problem.
In addition,  we prove lower bounds on the blowup rate for blowup solutions
of focusing GP-hierarchies in \cite{chpa2}.  

In the joint work \cite{chpatz1} with Tzirakis, 
we identify a conserved energy functional $E_1(\Gamma(t))=E_1(\Gamma_0)$ describing the
average energy per particle
(the precise definition is given in \eqref{eq-E1energy-def-0-1} below), 
and we prove virial identities for solutions of GP hierarchies.
In particular, we use these ingredients to prove that for $L^2$-critical
and supercritical focusing GP hierarchies, blowup occurs 
whenever $E_1(\Gamma_0)<0$ and the variance is finite.
We note that prior to  \cite{chpatz1}, no exact conserved energy functional on 
the level of the GP hierarchy was identified in any of the previous works,
including \cite{kiscst} and \cite{esy1,esy2}.

\subsection{Main results of this paper} 
We emphasize again that  our results in \cite{chpa2}, 
which are quoted in Theorem \ref{thm-localwp-TTstar-1-2} below, 
imply the local well-posedness of solutions for the spaces considered by Klainerman and Machedon,
\cite{klma}; those are of the form $\cH_\xi^\alpha$, under
constraints similar to  \eqref{intro-KMbound}.
However, they do not hold for the spaces considered by Erd\"os, Schlein and Yau, \cite{esy1,esy2}, 
%While the approach via space-time norms introduced by  Klainerman and Machedon allows for new and
%shorter proofs of uniqueness in spaces of the form $\cH_\xi^\alpha$ under
%the constraint \eqref{intro-KMbound}, \cite{klma,chpa2}, these  
which are of the form $\frH_\xi^1$ introduced below.
%In particular, our results in \cite{chpa2} do not imply  the uniqueness of solutions 
%proven by these authors in \cite{esy1,esy2} .
%One of the main results of the paper at hand is a new uniqueness proof that holds for the
%spaces $\frH_\xi^1$. 

To be more precise, we define the spaces
\eqn
	\frH_\xi^\alpha=\{ \, \Gamma\in\Gspace \,  | \, \| \, \Gamma \, \|_{\frH_\xi^\alpha} <\infty \, \}
\eeqn
with 
\eqn
	\| \, \Gamma \, \|_{\frH_\xi^\alpha} \, := \,
	\sum_{k\in\N} \xi^k \, \| \, \gamma^{(k)} \, \|_{\frh^\alpha} \,,
\eeqn
where
\eqn\label{eq-gamma-norm-def-0}
	\| \gamma^{(k)} \|_{\frh^\alpha} & := &  \tr( \, | S^{(k,\alpha)} \gamma^{(k)} | \, )  \,.
\eeqn 
 
For the {\em existence of solutions} $\Gamma(t)\in\frH_\xi^1$, we invoke the above noted
results of Erd\"os, Schlein and Yau, \cite{esy1,esy2}, where a global in time solution
is constructed from the solution of an $N$-body Schrodinger equation and the associated
BBGKY hierarchy.
We note that
expressed in our notation, the a priori energy bound
\eqref{eq-ESY-apriori-enbd-0} implies that for any $0<\xi<C_0^{-1}$, one has
\eqn 
	\| \, \Gamma(t) \, \|_{\frH_\xi^1} \, < \, \sum_{k\geq1} (C_0\xi)^k \, < \, \infty 
	\; \; \; \; , \; \; \; \; t\in\R \,.
\eeqn
Hence, the solutions of the cubic defocusing GP hierarchy derived by Erd\"os,
Schlein and Yau in \cite{esy1,esy2} are contained in the spaces $\frH_\xi^1$ considered
in this paper. 
Similarly, solutions of the quintic GP hierarchy obtained in \cite{chpa} (along the 
lines described in Section \ref{subsec-exsol-GP-0}) are   contained in
$\frH_\xi^1$.

The main results proven in this paper are:
\begin{enumerate} 

\item
We introduce
a new family of {\em  higher order energy functionals}, 
generalizing those found in \cite{chpatz1}, and prove that they are
conserved for solutions of the GP hierarchy (see Section \ref{sec-highordenergy-1}). 
\\
\item
We introduce a generalization
of Sobolev and Gagliardo-Nirenberg inequalities on the level of marginal density matrices
(see Section \ref{sec-Sobolev-1}). 
\\
\item
We prove a priori energy bounds on positive semidefinite solutions $\Gamma(t)\in\frH_\xi^1$
for defocusing, energy-subcritical GP hierarchies (see Subsection \ref{sec-globalwp-1}),
and for focusing, $L^2$-subcritical GP hierarchies (see Subsection
\ref{sec-globwp-L2subcrit-1}). 
Our argument is based on the conservation of the higher order energy functionals,
and employs the above noted Sobolev inequalities for marginal density matrices
(see Section \ref{sec-Sobolev-1}). 
%\\
%\item
%As a corollary, we prove 
%a priori bounds on solutions $\Gamma(t)\in\frH_\xi^1$
%of $p$-GP hierarchies that are defocusing and energy subcritical, or
%(de)focusing and  $L^2$-subcritical, without the factorization condition
%used in  \cite{esy1,esy2}.
%the uniqueness of solutions  $\Gamma(t)\in\frH_\xi^1$
%of $p$-GP hierarchies that are defocusing and energy subcritical, or
%(de)focusing and  $L^2$-subcritical.
%As a consequence, we provide a new proof of the uniqueness result of \cite{esy1,esy2}.
%In particular, our approach 
%does not invoke any Feynman graph expansions or Duhamel expansions at all
%(see Section \ref{sec-newproofuniq}).
\\
\item 
Finally, we  use the
the higher order energy functionals in order to enhance local to global wellposedness for 
solutions in the spaces $\cH_\xi^1$ constructed in \cite{chpa2}, for initial data in $\frH_\xi^1$ 
(see Section \ref{sec-globalwp-1-1}).  
%\\
%\item
%As an immediate consequence of the above results, 
%we find that the solutions found by Erd\"os, Schlein and Yau in \cite{esy1,esy2} and those obtained 
%in \cite{chpa2}
%for the same GP hierarchy as in \cite{esy1,esy2}, coincide for initial data in $\frH_\xi^1$.
\end{enumerate}
  
We note that our a priori upper bounds on the norm $\|\Gamma\|_{\frH_\xi^1}$ hold for 
positive semidefinite $\Gamma$.
The condition of positive definiteness is physically meaningful
because the components of $\Gamma$ are interpreted as density matrices.
An obvious example of such $\Gamma$
is given by arbitrary linear superpositions of factorized states with positive coefficients,
$\Gamma_0=\sum \mu_j \Gamma_{\phi_j}$, $\mu_j>0$, and 
$\Gamma_{\phi_j}=((|\phi_j\ket\bra\phi_j|)^{\otimes n})_{n\in\N}$, $\phi_j\in H^1(\R^d)$. 
Also, $\Gamma(t)$ obtained from the $N\rightarrow\infty$ limit of the BBGKY hierarchy
of an $N$-body Schr\"{o}dinger equation is generally expected to be positive semidefinite.
However,  the question whether positive semidefiniteness is generally preserved
by the flow of the GP hierarchy or not is not well understood, as far as we know,
and we leave a more systematic 
study of this issue to future work.

\newpage

\section{Definition of the model}
\label{sec-modeldef-1}

In this section, we introduce the mathematical model analyzed in 
this paper. Most notations and definitions are adopted from \cite{chpa2},
where we refer for motivations and more details.
 
\subsection{The spaces of marginal density matrices}
We consider the space introduced in \cite{chpa2} 
\eqn \nonumber 
	\Gspace \, := \, \bigoplus_{k=1}^\infty L^2(\R^{dk}\times\R^{dk})  
\eeqn
of sequences of density matrices
\eqn \nonumber 
	\Gamma \, := \, (\, \gamma^{(k)} \, )_{k\in\N}
\eeqn
where $\gamma^{(k)}\geq0$, $\tr\gamma^{(k)} =1$,
and where every $\gamma^{(k)}(\ux_k,\ux_k')$ is symmetric in all components of $\ux_k$,
and in all components of $\ux_k'$, respectively, i.e. 
\begin{equation}\label{symmetry}
	\gamma^{(k)}(x_{\pi (1)}, ...,x_{\pi (k)};x_{\pi'( 1)}^{\prime},
	 ...,x_{\pi'(k)}^{\prime})=\gamma^{(k)}( 	x_1, ...,x_{k};x_{1}^{\prime}, ...,x_{k}^{\prime})
\end{equation}
\\
holds for all $\pi,\pi'\in S_k$. 

Throughout this paper, we will denote vectors $(x_1, \cdots, x_k)$ by 
$\ux_k$ and vectors $(x'_1, \cdots, x'_k)$ by $\ux'_k$.
 
The $k$-particle marginals are assumed to be hermitean,
\begin{equation}\label{conj}
	\gamma^{(k)}(\ux_k;\ux_k')=\overline{\gamma^{(k)}(\ux_k';\ux_k) }.
\end{equation}
We call $\Gamma=(\gamma^{(k)})_{k\in\N}$ admissible if  
$\gamma^{(k)}=\tr_{k+1} \gamma^{(k+1)}$, that is,
\eqn 
	\gamma^{(k)}(\ux_k;\ux_k') 
	\, = \, \gamma^{(k+1)}(\ux_{k},x_{k+1};\ux_k',x_{k+1}) 
	\nonumber
\eeqn  
for all $k\in\N$.

Let $0<\xi<1$.   In \cite{chpa2}, we introduced the Hilbert-Schmidt type 
generalized Sobolev spaces  of sequences of marginal density matrices
\eqn\label{eq-cHalpha-def-1-1-1} 
 	\cH_\xi^\alpha \, := \, \Big\{ \, \Gamma \, \in \, 
	\Gspace \, \Big| \, \|\Gamma\|_{\cH_\xi^\alpha} < \, \infty \, \Big\} \,,
\eeqn
where
\eqn \nonumber 
	\|\Gamma\|_{\cH_\xi^\alpha} \, = \, \sum_{k=1}^\infty \xi^{ k} 
	\| \,  \gamma^{(k)} \, \|_{H^\alpha_k} \,,
\eeqn
with
\eqn\label{eq-gamma-norm-def-1}
	\| \gamma^{(k)} \|_{H^\alpha_k} \, := \, 
	\big( \, \tr( \, | S^{(k,\alpha)}  \gamma^{(k)}|^2 \, ) \, \big)^{\frac12}
\eeqn 
and $S^{(k,\alpha)}:=\prod_{j=1}^k\bra\nabla_{x_j}\ket^\alpha\bra\nabla_{x_j'}\ket^\alpha$.
They also correspond to the spaces of solutions studied in \cite{klma}.

In contrast, we also define the ${\mathcal L}^1$-Schatten class type generalized Sobolev spaces
\eqn\label{eq-frHalpha-def-1} 
 	\frH_\xi^\alpha \, := \, \Big\{ \, \Gamma \, \in \, \Gspace \, \Big| \, \|\Gamma\|_{\frH_\xi^\alpha} < \, \infty \, \Big\} \,,
\eeqn
where
\eqn \nonumber 
	\|\Gamma\|_{\frH_\xi^\alpha} \, = \, \sum_{k=1}^\infty \xi^{ k} 
	\| \,  \gamma^{(k)} \, \|_{\frh^\alpha_k} \,,
\eeqn
with
\eqn\label{eq-gamma-norm-def-2}
	\| \gamma^{(k)} \|_{\frh^\alpha_k} & := & \tr( \, | S^{(k,\alpha)}  \gamma^{(k)}| ) \, 
\eeqn 
that correspond to the spaces of solutions studied in \cite{esy1,esy2}.

\subsection{The Gross-Pitaevskii (GP) hierarchy} 
To unify the notation for cubic and quintic GP hierarchies,  
we introduced the notion of $p$-GP hierarchy  in \cite{chpa2}, given as follows. 
Let $p\in\{2,4\}$. Then, the $p$-GP  hierarchy is given by
\eqn \label{eq-def-GP}
	i\partial_t \gamma^{(k)} \, = \, \sum_{j=1}^k [-\Delta_{x_j},\gamma^{(k)}]   
	\, + \,  \mu B_{k+\frac p2} \gamma^{(k+\frac p2)}
\eeqn
for $k\in\N$. Here,
\eqn \label{eq-def-b}
	B_{k+\frac p2}\gamma^{(k+\frac{p}{2})}
	\, = \, B^+_{k+\frac p2}\gamma^{(k+\frac{p}{2})}
        - B^-_{k+\frac p2}\gamma^{(k+\frac{p}{2})} \, ,
\eeqn
where 
$$ B^+_{k+\frac p2}\gamma^{(k+\frac{p}{2})}
   = \sum_{j=1}^k B^+_{j;k+1,\dots,k+\frac p2}\gamma^{(k+\frac{p}{2})},
$$ 
and 
$$ B^-_{k+\frac p2}\gamma^{(k+\frac{p}{2})}
   = \sum_{j=1}^k B^-_{j;k+1,\dots,k+\frac p2}\gamma^{(k+\frac{p}{2})},
$$                  
with 
\begin{align*} 
& \left(B^+_{j;k+1,\dots,k+\frac p2}\gamma^{(k+\frac{p}{2})}\right)
(t,x_1,\dots,x_k;x_1',\dots,x_k') \\
& \quad \quad = \int dx_{k+1}\cdots dx_{k+\frac p2} dx_{k+1}'\cdots dx_{k+\frac p2}' \\
& \quad\quad\quad\quad 
	\prod_{\ell=k+1}^{k+\frac p2} \delta(x_j-x_{\ell})\delta(x_j-x_{\ell}' )
        \gamma^{(k+\frac p2)}(t,x_1,\dots,x_{k+\frac p2};x_1',\dots,x_{k+\frac p2}'),
\end{align*} 
and 
\begin{align*} 
& \left(B^-_{j;k+1,\dots,k+\frac p2}\gamma^{(k+\frac{p}{2})}\right)
(t,x_1,\dots,x_k;x_1',\dots,x_k') \\
& \quad \quad = \int dx_{k+1}\cdots dx_{k+\frac p2} dx_{k+1}'\cdots dx_{k+\frac p2}' \\
& \quad\quad\quad\quad 
	\prod_{\ell=k+1}^{k+\frac p2} \delta(x'_j-x_{\ell})\delta(x'_j-x_{\ell}' )
        \gamma^{(k+\frac p2)}(t,x_1,\dots,x_{k+\frac p2};x_1',\dots,x_{k+\frac p2}').
\end{align*} 
%Moreover, we let
%\eqn
%   B^{\pm}_{j;k+1,\dots,k+\frac p2}\gamma^{(k+\frac{p}{2})} 
%  \, := \,  B^+_{j;k+1,\dots,k+\frac p2}\gamma^{(k+\frac{p}{2})}
%      \, - \, B^-_{j;k+1,\dots,k+\frac p2}\gamma^{(k+\frac{p}{2})} \,.
%\eeqn
The operator $B_{k+\frac p2}\gamma^{(k+\frac{p}{2})}$
accounts for  $\frac p2+1$-body interactions between the Bose particles.
We note that for factorized solutions, the corresponding 1-particle wave function satisfies the
$p$-NLS $i\partial_t\phi=-\Delta\phi+\mu|\phi|^p\phi$.

Following our conventions in \cite{chpa2, chpatz1}, we refer to (\ref{eq-def-GP}) 
as the {\em cubic GP hierarchy} if $p=2$,
and as the {\em quintic GP hierarchy} if $p=4$. 
Moreover, we denote the $L^2$-critical exponent by $p_{L^2} = \frac 4d$,
and refer to  (\ref{eq-def-GP}) as a:
\begin{itemize}
\item {\em $L^2$-critical GP hierarchy} if $p=p_{L^2}$.
\item {\em $L^2$-subcritical GP hierarchy} if $p < p_{L^2}$.
\item {\em $L^2$-supercritical GP hierarchy} if $p > p_{L^2}$.
\end{itemize}
In an analogous manner, we use the notion of energy-critical (respectively, energy-subcritical and 
energy-supercritical GP hierarchies) if $p = p_{H^1}$ (respectively, $p < p_{H^1}$ and 
$p > p_{H^1}$), where $p_{H^1} = \frac{4}{d-2}$.    
Moreover, we respectively refer to the cases $\mu=1$ or $\mu=-1$ as
defocusing or focusing GP hierarchies. 

For notational brevity, we introduced the following compact notation for the $p$-GP 
hierarchy in \cite{chpa2},
\eqn \label{chpa2-pGP}
        i\partial_t \Gamma \, + \, \opDelta_\pm \Gamma & = & \mu \opB \Gamma 
        \nonumber\\
        \Gamma(0) &=& \Gamma_0 \,,
\eeqn 
where
$$
	\opDelta_\pm \Gamma \, := \, ( \, \Delta^{(k)}_\pm \gamma^{(k)} \, )_{k\in\N}
        \; \; \; \; \mbox{ with } \; \; \; \; \Delta_{\pm}^{(k)} \, = \, \Delta_{\ux_k} - \Delta_{\ux'_k},
$$
and 
\eqn \label{chpa2-B} 
	\opB \Gamma \, := \, ( \, B_{k+\frac{p}{2}} \gamma^{(k+\frac{p}{2})} \, )_{k\in\N} \,.
\eeqn 
Moreover, we will use the notation 
\begin{align*} 
	& \opB^+ \Gamma := \, ( \, B^+_{k+\frac{p}{2}} \gamma^{(k+\frac{p}{2})} \, )_{k\in\N}, 
        \nonumber \\  
	& \opB^- \Gamma := \, ( \, B^-_{k+\frac{p}{2}} \gamma^{(k+\frac{p}{2})} \, )_{k\in\N} \,.
\end{align*}
%We refer to \cite{chpa2} for more detailed explanations.

\section{Statement of the main Theorems}
\label{sec-mainthms-1}

In this section, we state the main results of this paper.
In Theorem \ref{thm-main-highordconsen-0}, we establish the conservation of novel
higher order energy functionals for solutions of $p$-GP hierarchies in $\frH_\xi^1$,
which are first introduced in this paper.
This result is used to prove a priori energy bounds on positive
semidefinite solutions for defocusing, energy subcritical
$p$-GP hierarchies in Theorem  \ref{thm-globsp-encons-0}, and for (de)focusing
$L^2$-subcritical $p$-GP hierarchies in Theorem \ref{thm-L2subcrit-0}.

We note that the local well-posedness of solutions in the space $\frH_\xi^1$ 
has so far been an open problem.
While the unconditional 
uniqueness (that is, without any requirement on the a priori boundedness
of some Strichartz norm) of solutions in $\frH_\xi^1$ has been proven in \cite{esy1,esy2},
existence of solutions is only established for factorized data. 
This is because in \cite{esy1,esy2},
solutions to the GP hierarchy have been derived only for initial data
for the $N$-body Schrodinger system that are
asymptotically factorizing.

Our local well-posedness result proven in \cite{chpa2} implies the  
well-posedness of solutions in $\cH_\xi^1$ with initial data in $\frH_\xi^1\subset\cH_\xi^1$, 
under the requirement that $\opB\Gamma\in L^2_t\cH^\alpha_\xi$ holds,
similar to the a priori space time bound introduced by Klainerman and Machedon in
\cite{klma}. Using the conserved, higher order energy functionals, we prove in this paper
that positive semidefinite 
solutions of this form remain in $\frH_\xi^1$, and can in fact be enhanced to global
solutions. This is demonstrated for the defocusing energy subcritical case, and for the $L^2$-subcritical
(de)focusing case. 

Our a priori upper bounds on the $\frH_\xi^1$-norm of $\Gamma$ hold for 
positive semidefinite $\Gamma$, as has been noted here and in the introduction.
Positive definiteness is physically meaningful because 
the components of $\Gamma$ are interpreted as density matrices.
An obvious example is given by arbitrary linear superpositions of factorized states 
with positive coefficients,
$\Gamma_0=\sum \mu_j \Gamma_{\phi_j}$, $\mu_j>0$, and 
$\Gamma_{\phi_j}=((|\phi_j\ket\bra\phi_j|)^{\otimes n})_{n\in\N}$, $\phi_j\in H^1(\R^d)$. 
Furthermore, $\Gamma(t)$ obtained from the $N\rightarrow\infty$ limit of the BBGKY hierarchy
of an $N$-body Schrodinger equation is generally expected to be positive semidefinite.
The question whether $\Gamma(t)$ is positive semidefinite whenever the same holds for $\Gamma_0$ 
is not well understood, as far as we know, and we leave a more systematic 
study of this issue to future work.

%Accordingly, in combination with the unconditional uniqueness
%of solutions in $\frH_\xi^1$ proven in \cite{esy1,esy2}, we conclude that 
%the Cauchy problem for the cubic, defocusing GP hierarchy
%in $\frH^1_\xi$ is unconditionally well-posed.

%As a consequence, we obtain the uniqueness
%of solutions of a large class of $p$-GP hierarchies in
%Theorem \ref{thm-main-uniqueness-0}, including the cases covered
%in \cite{esy1,esy2}. 
%: Energy subcritical, defocusing $p$-GP hierarchies, and $L^2$-subcritical focusing and 
% defocusing $p$-GP hierarchies.     
%, which is an improvement on our result in \cite{chpa2}.

%For the {\em existence of solutions} $\Gamma(t)\in\frH_\xi^1$, we refer to the work 
%of Erd\"os, Schlein and Yau, \cite{esy1,esy2}.
%As described in Section \ref{subsec-exsol-GP-0}, these authors
%derive a global in time solution of the $p$-GP hierarchy for $p=1$, $d=3$, $\mu=1$, 
%obtained from the weak subsequential 
%$N\rightarrow\infty$ limit
%of the BBGKY hierarchy associated to an $N$-body Schr\"{o}dinger equation 
%\eqref{ham1}
%in Gross-Pitaevskii scaling.

\begin{theorem}\label{thm-main-highordconsen-0}
Let $\Gamma(t)\in \frH_\xi^1$
be a solution of the $p$-GP hierarchy, for $t\in I\subseteq\R$ with $\{0\}\in I$, and some $0<\xi<1$.
Then, the infinite family of operators $(\cK^{(m)})_{m\in\N}$ presented in \eqref{eq-cKm-def-1} 
below define  an infinite sequence of linear functionals,
\eqn 
  \Gamma(t) \, \mapsto \, (\bra \, \cK^{(m)} \, \ket_{\Gamma(t)} )_{m\in\N} 
\eeqn  
where
\eqn 
  \bra \, \cK^{(m)} \, \ket_{\Gamma(t)} \, := \, \tr( \, \cK^{(m)} \, \gamma^{(m k_p)}(t) \, ) \,  ,
\eeqn
which we refer to as {\bf higher order energy functionals}.
Here, we recall that $\Gamma(t)=(\gamma^{(n)}(t))_{n\in\N}$ and $k_p=1+\frac p2$.
The higher order energy functionals have the following properties:
\begin{itemize}
\item
They are bounded, and in particular, there exists $0<\hat \xi<\xi$ such that
\eqn
	\sum_{m\in\N}\hat\xi^m \bra \cK^{(m)} \ket_{\Gamma(t)} 
	\, \leq \,  \|\Gamma(t)\|_{\frH_{\xi}^1}
\eeqn
holds, for all $t\in I$.
\item
The higher order energy functionals are conserved,
\eqn 
	\bra \, \cK^{(m)} \, \ket_{\Gamma(t)} \, = \, \bra \, \cK^{(m)} \, \ket_{\Gamma(0)}
\eeqn
for all $t\in I$, and all $m\in \N$.
\item
Assuming that the initial condition satisfies $\Gamma(0)\in\frH_{\xi'}^{1}$ for some $0<\xi'<1$,
and that 
\eqn\label{eq-xixiprime-Sob-0}
	\xi \, \leq \,     \big( \, 1+\frac{2}{p+2}\CSob (d,p) \, \big)^{-\frac{1}{k_p}} \, \xi' \,,
\eeqn
where the constant $\CSob(d,p)$ is as in Theorem \ref{eq-Sobmultidim-1} below,
the a priori bound 
\eqn\label{eq-Km-highenfunct-def-0}
	\sum_{m\in\N}(2\xi)^m \bra \cK^{(m)} \ket_{\Gamma(t)} 
	\, \leq \,  \|\Gamma_0\|_{\frH_{\xi'}^1}
\eeqn
is satisfied for  $p < \frac{4}{d-2}$ and $|\mu|\leq1$ (focusing and defocusing hierarchies).
\end{itemize}
\end{theorem}

%The key improvement of this local well-posedness result
%over the one established in \cite{chpa2} consists of the fact that the 
%initial condition and the solution are in the same space $\frH_{\xi}^\alpha$.
%In \cite{chpa2}, the initial data $\Gamma_0$ was required to belong to $\frH_{\xi_1}^\alpha$
%for some $\xi_1>0$, while the solution $\Gamma(t)$ was shown to belong to
%$\frH_{\xi_2}^\alpha$, for some $0<\xi_2<\xi_1$. For the proof of Theorem \ref{thm-localwp-TTstar-1},
%we refer to \cite{chpa3}.
%\\

Accordingly, the higher order energy functionals are bounded, and for sufficiently small $\xi>0$, the
associated power series \eqref{eq-Km-highenfunct-def-0} is bounded by the initial data of the 
solution $\Gamma(t)\in \frH_\xi^1$, due to their time invariance.

On the other hand, the higher order energy functionals provide us with a priori bounds on the
$\frH_\xi^1$ norm of the solution itself. In case of defocusing $p$-GP hierarchies, we find the
following result.

\begin{theorem}
\label{thm-globsp-encons-0}
Assume that $\mu=+1$ (defocusing $p$-GP hierarchy), 
$p <  \frac{4}{d-2}$,  
and that $\Gamma(t)\in\frH^{1}_\xi$ for $t\in  [0,T]$, 
is a positive semidefinite solution of the $p$-GP hierarchy with initial data  $\Gamma_0 \in \frH_{\xi'}^1$  
for $\xi$, $\xi'$ as in \eqref{eq-xixiprime-Sob-0}. Then, the a priori bound 
\eqn 
	\|\Gamma(t)\|_{\frH_{\xi}^1} & \leq &\sum_{m\in\N}(2\xi)^m \bra \cK^{(m)} \ket_{\Gamma(t)}  
	\nonumber\\
	& = &
	\sum_{m\in\N}(2\xi)^m \bra \cK^{(m)} \ket_{\Gamma_0}  \, \leq \,  \|\Gamma_0\|_{\frH_{\xi'}^1}
\eeqn
is satisfied for all $t\in [0,T]$.
\end{theorem}

For focusing, $L^2$-subcritical $p$-GP hierarchies, we obtain a similar result, provided that the interaction is
not too large.

\begin{theorem}\label{thm-L2subcrit-0}
Let $p<p_{L^2}=\frac4d$ ($L^2$ subcritical).
Moreover, let $\alpha>\alpha_0:=\frac{(k_p-1)d}{2k_p}$ and $\alpha k_p <1$,
where $k_p=1+\frac p2$, and $\alpha<1$.  Let 
\eqn 
D: = D(\alpha, p, d, |\mu|) =   \left( 1 - |\mu|  \frac{C_0(\alpha)}{1-4^{-(1-\alpha k_p)} } \right),
\label{eq-defD} 
\eeqn
where $C_0(\alpha)$ is characterized in \eqref{eq-Bgamma-GaglNir-1}.

Assume that  $\Gamma(t)\in\frH_\xi^{1}$ 
is a positive semidefinite solution of the focusing ($\mu<0$) $p$-GP hierarchy 
for $t\in I$,
with initial data $\Gamma(0)=\Gamma_0\in\frH_{\xi'}^{1}$,  
where 
\eqn\label{eq-xixiprime-Sob-0-5}
	\xi \, \leq \,   \frac{1}{D} \,  \big( \, 1+\frac{2}{p+2}\CSob (d,p) \, \big)^{-\frac{1}{k_p}} \, \xi' \,.
\eeqn 
If $\mu<0$ is such that  
\eqn \label{eq-mu-L2sub-uppbd-0}
  |\mu| \, < \,  \frac{ 1-4^{-(1-\alpha k_p)}  }{ C_0(\alpha)  } \,,
\eeqn 
then the a priori bound
\eqn
  \| \, \Gamma(t) \, \|_{\frH_{\xi}^1}  
  \, & \leq & \, \sum_{m=1}^\infty \, (2 D \, \xi)^m \, \bra \, \cK^{(m)} \, \ket_{\Gamma(t)} 
  \label{eq-H1-aprioribd-firstline} \\
  \, & =  & \, \sum_{m=1}^\infty \, (2 D \, \xi)^m \, \bra \, \cK^{(m)} \, \ket_{\Gamma_0}
  \\\label{eq-5xiKm-apriori-0-0}
  &\leq& \|\Gamma_0\|_{\frH_{\xi'}^1}  
\label{eq-H1-aprioribd-2}
\eeqn
holds for all $t\in I$.  
 
\end{theorem}

%In \cite{esy1,esy2}, Erd\"os, Schlein and Yau prove the unconditional uniqueness of solutions
%in $L_t^\infty\frH_\xi^1$ for the cubic, defocusing GP hierarchy in $d=3$. 

%\begin{theorem}\label{thm-main-uniqueness-0} 
%(Erd\"os, Schlein and Yau, \cite{esy1,esy2})
%Let $\Gamma_1(t)$, $\Gamma_2(t)\in \frH_\xi^{1}$, for $t\in [0,T]$,
%be two solutions of the defocusing, cubic GP hierarchy in $d=3$,
%\eqn 
%   i\partial_t \Gamma \, + \, \opDelta_\pm \Gamma & = & \mu  \opB \Gamma 
%\eeqn 
%with identical initial data,  $\Gamma_1(0)=\Gamma_2(0)=\Gamma_0\in\frH_{\xi'}^{1}$, for 
%$\xi$, $\xi'$ satisfying
%\eqref{eq-xixiprime-Sob-0}, respectively \eqref{eq-xixiprime-Sob-0-5}.
%Then, 
%%if
%\begin{itemize}
 %\item $p<\frac{4}{d-2}$ (energy subcritical), and $\mu=+1$ (defocusing), respectively
 %\item $p<\frac4d$ ($L^2$ subcritical), and $\mu<0$ (focusing) with   
 %$|\mu| \, < \,  \frac{ 1-4^{-(1-\alpha k_p)} }{ C_0(\alpha)  }$,
%%\end{itemize}
%it follows that $\Gamma_1(t)=\Gamma_2(t)$ for all $t\in [0,T]$.
%\end{theorem}

In \cite{chpa2}, we proved local well-posedness of solutions in the spaces 
$\Gamma\in L_t^\infty\cH_\xi^\alpha$ under the condition that 
$\opB\Gamma\in L^2_t\cH^\alpha_\xi$. 

\begin{theorem}
\label{thm-localwp-TTstar-1-2}
Let
\eqn\label{eq-alphaset-def-0-0} 
	\alpha \, \in \, \alphaset(d,p)
	\, := \,  \left\{
	\begin{array}{cc}
	(\frac12,\infty) & {\rm if} \; d=1 \\ 
	(\frac d2-\frac{1}{2(p-1)}, \infty) & {\rm if} \; d\geq2 \; {\rm and} \; (d,p)\neq(3,2)\\
	\big[1,\infty) & {\rm if} \; (d,p)=(3,2) \,.
	\end{array}
	\right.
\eeqn
Then, there exists a constant $0<\eta<1$ such that for $0<\xi\leq\eta \, \xi'\leq1$, 
there exists a constant $T_0(d,p,\xi,\xi')>0$ such that
the following holds.  Let $I:=[0,T]$ for $0<T<T_0(d,p,\xi,\xi')$.  
Then, there exists a unique solution
%there exists $0<\xi_2<\min\{1,\xi\}$ such that
$\Gamma\in L^\infty_{t\in I}\cH_{\xi}^\alpha$ of the $p$-GP hierarchy, with 
\eqn 
  \| \, \opB\Gamma \, \|_{ L^1_{t\in I}\cH_{\xi}^\alpha } \, < \, 
  C(T,\xi,\xi',d,p) \, \|\Gamma_0\|_{\cH_{\xi'}^\alpha} \,,
\eeqn
in the space 
\eqn\label{eq-Wspace-def-1-2}
	{\mathcal W}^\alpha(I,\xi) \, = \, \{ \, \Gamma\in L^\infty_{t\in I}\cH_{\xi}^\alpha \, | \,
	\opB^{+}\Gamma \, , \, \opB^{-}\Gamma \in  L^2_{t\in I}\cH_{\xi}^\alpha \, \}
\eeqn
for the initial condition $\Gamma(0)=\Gamma_0\in\cH_{\xi'}^\alpha$. 
\end{theorem}

% 
% The key improvement of this local well-posedness result
% over the one established in \cite{chpa2} consists of the fact that the 
% initial condition and the solution are in the same space $\cH_{\xi}^\alpha$.
% In \cite{chpa2}, the initial data $\Gamma_0$ was required to belong to $\cH_{\xi_1}^\alpha$
% for some $\xi_1>0$, while the solution $\Gamma(t)$ was shown to belong to
% $\cH_{\xi_2}^\alpha$, for some $0<\xi_2<\xi_1$. For the proof of Theorem \ref{thm-localwp-TTstar-1},
% we refer to \cite{chpa3}.
% \\

We note that the presence of two different energy scales $\xi,\xi'$ has the 
following interpretation on the level of the NLS. Let $R_0:=(\xi')^{-1/2}$ and $R_1:=\xi^{-1/2}$.
Then, the local well-posedness result in Theorem \ref{thm-localwp-TTstar-1-2}, applied to
factorized initial data $\Gamma_0=\Gamma_{\phi_0}$, is equivalent to the following statement:
For $\|\phi_0\|_{H^1(\R^n)}<R_0$, there exists a unique solution $\|\phi\|_{L^\infty_{t\in I}H^1(\R^n)}<R_1$,
with $R_1>R_0$,
in the space 
$$\{\phi \in L^\infty_{t\in I}H^1(\R^n) \, | \, \||\phi|^p\phi\|_{L^2_tH^1}<\infty\}\,.$$
This version of local well-posedness, specified  for  balls $B_{R_0}(0),B_{R_1}(0)\subset H^1(\R^n)$,
contains the less specific formulation of local well-posedness where only finiteness is required,
$\|\phi_0\|_{H^1(\R^n)}<\infty$ and $\|\phi\|_{L^\infty_{t\in I}H^1(\R^n)}<\infty$.
%There appear to be some misunderstandings in the literature as to the interpretation
%of our results, \cite{chz2}.

In this paper we enhance local to global wellposedness for 
solutions in the spaces $\cH_\xi^1$ constructed in \cite{chpa2}, for initial 
data in $\frH_{\xi'}^1\subset\cH_{\xi'}^1$ with $\xi<\xi'$,
provided that the solution is positive semidefinite.  
The proof is again based on the use of
the higher order energy functionals.

\begin{theorem}
\label{thm-globsp-encons-1-2}
Assume one of the following two cases:
\begin{itemize}
 \item  Energy subcritical, defocusing $p$-GP hierarchy with $p<\frac{4}{d-2}$ and  $\mu=+1$. Moreover, $\xi$, $\xi'$ satisfy 
    \eqref{eq-xixiprime-Sob-0}.
 \item  $L^2$ subcritical, focusing $p$-GP hierarchy with  $p<\frac4d$ and $\mu<0$ satisfying  
 \eqref{eq-mu-L2sub-uppbd-0}. In addition, $\xi$, $\xi'$ satisfy 
    \eqref{eq-xixiprime-Sob-0-5}.
\end{itemize}
Then, there exists $T>0$ such that for $I_j:=[jT,(j+1)T]$, with $j\in\Z$, there exists  
a unique global solution $\Gamma\in \cup_{j\in\Z}{\mathcal W}^1(I_j,\xi)$ of the $p$-GP hierarchy
with initial condition $\Gamma(0)=\Gamma_0\in\cH_{\xi'}^1$, satisfying
\eqn
	\|\Gamma(t)\|_{\cH_{\xi}^1} \, \leq \, \|\Gamma(t)\|_{\frH_{\xi}^1} \, \leq \, C  \|\Gamma_0\|_{\frH_{\xi'}^1} 
\eeqn
for all $t\in\R$, if $\Gamma(t)$ is positive semidefinite for all $t\in I_j$, $j\in\Z$.
\end{theorem}
 
%As an immediate consequence of Theorem \ref{thm-main-uniqueness-0},
%Theorem \ref{thm-localwp-TTstar-1-2}
%and Theorem \ref{thm-globsp-encons-1-2},
%we obtain the unconditional well-posedness of the Cauchy problem for the
%cubic, defocusing GP hierarchy in $d=3$.

%\begin{corollary}
%Let $\Gamma_0\in  \frH_{\xi'}^1$ satisfy the conditions of symmetry, admissibility, and positive
%semidefiniteness.
%Then, there exists a unique global solution $\Gamma\in L^\infty_t\frH_\xi^1$ 
%of the cubic defocusing GP hierarchy in $d=3$ with $\Gamma(0)=\Gamma_0$, for 
%$\xi$, $\xi'$ satisfying \eqref{eq-xixiprime-Sob-0}. 
%\end{corollary}

%We note again that the condition of cubic, defocusing GP hierarchy in $d=3$ appears here because the
%unconditional uniqueness of solutions has been established by \cite{esy1,esy2} only
%for this case.

\section{Higher order energy conservation}
\label{sec-highordenergy-1}

In this section, we introduce a higher order generalization of the energy
functional introduced in \cite{chpatz1}. We prove that it is a conserved quantity
for solutions of the $p$-GP hierarchy. 
As a main application, this conserved quantity will be used
%to obtain the new proof of uniqueness of solutions to the GP hierarchy which invokes
%neither Feynman graphs nor the board game argument
%(see Section \ref{sec-newproofuniq} for details). 
%Moreover, we use higher order energy functional 
to enhance local well-posedness (obtained in  \cite{chpa2}) to global well-posedness
for  certain defocusing GP-hierarchies in the spaces $\cH_\xi^1$,
for initial data in $\frH_\xi^1$ (see Section \ref{sec-globalwp-1-1} for details).

Let 
\eqn
  k_p \, := \, 1 \, + \, \frac p2 \,.
\eeqn
We define the operators
\eqn  
  K_\ell 
  \, := \,  \frac{1}{2} \, ( 1 \, - \, \Delta_{x_\ell}) \, \tr_{\ell+1,\dots,\ell+\frac p2}  
  \, + \, \frac{\mu}{p+2} \,  B_{\ell;\ell+1,\dots,\ell+\frac p2}^+
  \nonumber
\eeqn
for $\ell\in\N$. This operator
is related to the average energy per particle $E_1(\Gamma)$ (introduced in \cite{chpatz1})
through
\eqn\label{eq-E1energy-def-0-1}
  \lefteqn{
  	\frac12+E_1(\Gamma) 
	}
	\nonumber\\
	& = & \tr_{1, \dots, \ell, \ell+k, \cdots, j} K_{\ell} \gamma^{(j)} 
  \nonumber\\
  & = &
  \frac{1}{2} \, + \, \frac{1}{2} \, \tr_1(-\Delta_x\gamma^{(1)}) 
  \, + \, \frac{\mu}{p+2}\int dx \, \gamma^{(k_p)}(x,\dots,x;x,\dots,x) \,,
  \; \; \;
\eeqn 
using the admissibility of $\Gamma=(\gamma^{j})_{j\in\N}$,
see also \cite{chpatz1}.

% As an example, for $p=2$ (cubic GP hierarchy), $\ell=2$ and 
% $\gamma^{(j)}$ with $j=5 \geq \ell+k_p=4$, we have
% \eqn 
%   \lefteqn{
%   (H^{(1)}_2\gamma^{(5)})(\utx_{2};\utx_{2}')
%   \, = \,  \tr_{j+1,\dots,j+\frac p2}( \, -\Delta_{x_\ell}\gamma^{(j)} \, )(\utx_n;\utx_n')
%   }
%   \\
%   &&
%   \, + \, \frac{\mu}{p+2} \, \gamma^{(n+\frac p2)}
%   (\ux_{\ell-1}, x_\ell,\dots,x_\ell,x_{\ell+\frac p2+1},\dots,x_j ;
%   \ux_{\ell}',x_\ell,\dots,x_\ell,x_{\ell+\frac p2+1}',\dots,x_j')
%   \nonumber
% \eeqn
% We define operators 
% \eqn 
%   \lefteqn{
%   (H^{(1)}_\ell\gamma^{(j)})(\ux_{j-\frac p2};\ux_{j-\frac p2}')
%   \, := \,  \tr_{j+1,\dots,j+\frac p2}( \, -\Delta_{x_\ell}\gamma^{(j)} \, )(\ux_n;\ux_n')
%   }
%   \\
%   &&
%   \, + \, \frac{\mu}{p+2} \, \gamma^{(n+\frac p2)}
%   (\ux_{\ell-1}, x_\ell,\dots,x_\ell,x_{\ell+\frac p2+1},\dots,x_j ;
%   \ux_{\ell}',x_\ell,\dots,x_\ell,x_{\ell+\frac p2+1}',\dots,x_j')
%   \nonumber
% \eeqn
% for $j > \ell+\frac p2$. The superscript ``$(1)$" accounts for the 
% fact that 
% \eqn 
%   \tr( \, H^{(1)}_\ell\gamma^{(j)} \, ) 
%   \, = \, \tr( \, H^{(1)}_1\gamma^{(1+\frac p2)} \, ) 
% \eeqn
% is the conserved one-particle energy (the rhs equals the lhs due to 
% the admissibility of  $\gamma^{(j)}$ and its symmetry with respect
% to the components of $\ux_j,\ux_j'$).

Moreover, we introduce the operator
\eqn\label{eq-cKm-def-1}
  \cK^{(m)} & := & K_{1} 
  K_{k_p + 1}  \cdots K_{(m-1) k_p + 1}
\eeqn
where the $m$ factors are mutually commuting, in the sense that 
\eqn
	K_{j k_p+1}K_{j' k_p+1}\gamma^{(j)} 
 	\, = \, K_{j' k_p+1} K_{j k_p+1}\gamma^{(j)} 
\eeqn 
holds for $0\leq j\neq j'\leq m-1$.

We may now give the precise statement of our main result that provides the
conservation of higher energy functional for solutions of the $p$-GP hierarchy.  
\begin{theorem}\label{thm-Enconserv-1}
Assume that  $\Gamma=(\gamma^{(j)})\in\frH_\xi^1$ is admissible and
solves the $p$-GP hierarchy.
%, for initial data $\Gamma_0\in\frH_{\xi'}^1$ where $\xi$, $\xi'$
%are as in \eqref{eq-xixiprime-Sob-0}. 
%Let $m\in\N$. 
Then, for all $m\in\N$, the {\bf higher order energy functionals}
\eqn
  \bra \cK^{(m)} \ket_{\Gamma(t)} \, := \, \tr_{1,k_p + 1 ,2k_p + 1,\dots,(m-1)k_p + 1} 
  ( \, \cK^{(m)} \, \gamma^{(m k_p)}(t) \, ) 
\eeqn
are bounded, and are conserved quantities, i.e. 
\eqn\label{eq-cKcons-id-1}
  \partial_t\bra \cK^{(m)} \ket_{\Gamma(t)} \, = \, 0 \,,
\eeqn 
for every $m\in\N$.
\end{theorem}

\begin{remark} 
As is shown in Theorem \ref{thm-bd-foc-defoc} below,  
 $\bra \cK^{(m)} \ket_{\Gamma(t)}$ are bounded, and in particular, there exists $0<\hat \xi<\xi$ such that
\eqn \label{eq-Kxi-Gamma-aprioribd-0-0} 
	\sum_{m\in\N}\hat\xi^m \bra \cK^{(m)} \ket_{\Gamma(t)} 
	\, \leq \,  \|\Gamma(t)\|_{\frH_{\xi}^1}
\eeqn
holds, for all $t\in I$.
% $\bra \cK^{(m)} \ket_{\Gamma(t)}$ satisfy the a priori bounds
 %\eqn\label{eq-Kxi-Gamma-aprioribd-0-0} 
%	\sum_{m\in\N}(2\xi)^m \bra \cK^{(m)} \ket_{\Gamma(t)} 
%	\, \leq \,  \|\Gamma(t)\|_{\frH_{\xi'}^1}
%\eeqn
%for  $\xi$, $\xi'$ as in \eqref{eq-xixiprime-Sob-0},
%where either  $\mu>0$ and $p < \frac{4}{d-2}$  (defocusing $p$-GP hierarchy),  or
%$\mu<0$ and $p<p_{L^2}=\frac4d$ (focusing $L^2$ subcritical GP hierarchy).
Hence, the conserved quantities  $\bra \cK^{(m)} \ket_{\Gamma(t)}$ are well
defined for solutions of the GP hierarchy $\Gamma=(\gamma^{(j)})\in\frH_{\xi}^1$.
\end{remark} 

\begin{remark} 
We note that in the definition of $ \bra \cK^{(m)} \ket_{\Gamma(t)} $,
we may replace $\gamma^{(m k_p)}$ by any $\gamma^{(j)}$ with $j\geq m k_p$,
and would still obtain the same value of $\bra \cK^{(m)} \ket_{\Gamma(t)}$.
\end{remark} 

%Moreover, $ \bra \cK^{(m)} \ket_{\Gamma(t)} $ is a conserved quantity, 
%\eqn\label{eq-cKcons-id-1}
 % \partial_t\bra \cK^{(m)} \ket_{\Gamma(t)} \, = \, 0 \,,
%\eeqn 
%for every $m\in\N$.
%\end{theorem}

\prf
To prove (\ref{eq-cKcons-id-1}), we note that
\eqn
  i\partial_t\gamma^{(m k_p)} & = & \sum_{\ell=1}^m
  \Big( \, h_\ell^\pm \gamma^{(m k_p)} \, + \, \mu \, b_\ell^\pm \gamma^{(mk_p+\frac{p}{2})} \, \Big)
\eeqn
where 
\eqn
  \lefteqn{
  h_\ell^\pm \gamma^{(m k_p)}(\ux_{m k_p};\ux_{m k_p}')
  }
  \\
  & := &
  - \, \sum_{j=(\ell - 1)k_p+1}^{\ell k_p}  (\Delta_{x_j}-\Delta_{x_j'}) 
  \, \gamma^{(m k_p)}(\ux_{m k_p};\ux_{m k_p}')
  \nonumber
\eeqn
and
\eqn
  \lefteqn{
\left(   b_\ell^\pm \gamma^{(m k_p+\frac{p}{2})}  \right) (\ux_{m k_p};\ux_{m k_p}')
  }
  \\
  & := &
  \sum_{j=(\ell - 1)k_p+1}^{\ell k_p} ( \, B_{j;m k_p+1,\dots,mk_p+\frac{p}{2}}^\pm
  \, \gamma^{(mk_p+\frac{p}{2})})(\ux_{m k_p};\ux_{m k_p}') \,.
  \nonumber
\eeqn
Accordingly,
\eqn\label{eq-partialt-cKm-def-1}
  \partial_t\bra \cK^{(m)} \ket_{\Gamma(t)} 
  \, = \,
  \sum_{\ell=1}^m \Big[ \,A (\ell;m)
  %\, + \, 
  %\mu \, A_b(\ell;m) 
  \, \Big] \,,
  \nonumber
\eeqn
where
\eqn\label{eq-Aellm-def-1}
  \lefteqn{
  A(\ell;m)
  }
  \\
  & := & \tr_{1,k_p + 1,2k_p + 1,\dots,(m-1)k_p + 1} 
  \Big( \, \cK^{(m)} \big( \, h_\ell^\pm \gamma^{(m k_p)} \, +\, \mu \, b_\ell^\pm \gamma^{(mk_p+\frac{p}{2})} \, \big) \, \Big) \,.
  \nonumber
\eeqn
%and
% \eqn
%   A_b(\ell;m) & := &
%   \tr_{1,k_p + 1,2k_p + 1,\dots,(m-1)k_p + 1}  ( \, \cK^{(m)} b_\ell^\pm \gamma^{((m+1)k_p)}\, ) \,.
% \eeqn
We claim that
\eqn 
  A(\ell;m)
%   \, + \, \mu \,  
%   A_b(\ell;m) 
  \, = \, 0
\eeqn 
for every $\ell\in\{1,\dots,m\}$.

To prove this, we first of all note that by symmetry of $\gamma^{(m k_p)}(\ux_{m k_p};\ux_{m k_p}')$
with respect to the components of $\ux_{m k_p}$ and $\ux_{m k_p}'$, it
suffices to assume that $\ell=1$. The other cases are similar.

Accordingly, letting $\ell=1$, we introduce the notations
\eqn
  \widetilde\gamma^{(k_p)}(\ux_{ k_p};\ux_{ k_p}')
  \, := \, \tr_{k_p+1,\dots,mk_p}( \, K_{k_p+1}\cdots K_{(m-1)k_p+1} \gamma^{(mk_p)} \, )
\eeqn
and
\eqn 
  \lefteqn{
  \widetilde\gamma^{(2k_p-1)}(\ux_{ k_p},\uy_{ k_p-1};\ux_{ k_p}',\uy_{ k_p-1}')
  }
  \\
  & := & \tr_{k_p+1,\dots,mk_p}( \, K_{k_p+1}\cdots K_{(m-1) k_p+1} \gamma^{(mk_p+\frac{p}{2})} \, )
  (\ux_{ k_p},\uy_{ k_p-1};\ux_{ k_p}',\uy_{ k_p-1}')
  \nonumber
\eeqn
where $y_i=x_{m k_p+i}$ and $y_i'=x_{m k_p+i}'$, for $i\in\{1,\dots,k_p-1\}$. 

In what follows, we will keep $m$ fixed, and omit it from the notation.
Given $\widetilde\gamma^{(j)}$ with $j\in\{k_p,2k_p-1\}$, we decompose  
\eqn
	\R^{jd} \, = \, \bigcup_{r\in\Z^{jd}} Q_r
\eeqn 
into disjoint cubes $Q_r$ obtained from translating the
unit cube $[0,1)^{jd}$ by  $r\in\Z^{jd}$.  We then define
\eqn 
  \widetilde\gamma^{(j)}_{r,r'} \, := \, P_{Q_r,Q_{r'}}(\widetilde\gamma^{(j)})
\eeqn
where $P_{Q_r,Q_{r'}}$ is the Fourier multiplication operator with symbol given by the characteristic
function of $Q_r\times Q_{r'}$. That is,
\eqn 
  \widehat{\widetilde\gamma^{(j)}_{r,r'}}(\uu_{j};\uu_{j}') \, = \, \chi_{Q_r}(\uu_{j})\chi_{Q_{r'}}(\uu_{j}')
  \widehat{\widetilde\gamma^{(j)}}(\uu_{j};\uu_{j}') 
\eeqn
for $r,r'\in \Z^{jd}$. 

Using the notation
\eqn
  K_1 \, = \, K_1^{(1)} \, + \,  K_1^{(2)} 
\eeqn
where
\eqn
  K_1^{(1)} \, := \, \frac{1}{2} \, ( 1 - \Delta_{x_1}) \, \tr_{2,\dots,k_p}
\eeqn
and
\eqn
  K_1^{(2)} \, := \, \frac{\mu}{p+2} \, B^+_{1;2,\dots,k_p} \,,
\eeqn
we consider
\eqn
  A_h^{(1)}(r,r') & := &\tr_1( \, K_1^{(1)} h_1^\pm \widetilde \gamma^{(k_p)}_{r,r'} \, )
  \nonumber\\
  & = &-
  \frac{1}{2}\tr_{1,2,\dots,k_p}( \, (1 - \Delta_{x_1}) 
  \, \sum_{j=1}^{k_p}(\Delta_{x_j}-\Delta_{x_j'}) \,
  \widetilde\gamma^{(k_p)}_{r,r'} \, )
  \nonumber\\
  & =& 
  \frac{1}{2} \int du_1 \dots du_{k_p} du'_1 \dots du'_{k_p} 
  \int dx_1 \dots dx_{k_p} dx'_1 \dots dx'_{k_p} 
  \nonumber\\ 
  && \quad \quad \delta(x_1 - x_1')  \cdots \delta(x_{k_p} - x'_{k_p}) \, 
  (1 + u_1^2) \, \sum_{j=1}^{k_p} \, \left( u_j^2 - (u'_j)^2 \right) 
  \nonumber\\ 
  && \quad \quad \quad \quad 
  \big( \, \prod_{l=1}^{k_p} e^{i(u_lx_l - u'_lx'_l)} \, \big) 
  \widehat{\widetilde\gamma^{(k_p)}_{r,r'} }(\uu_{ k_p};\uu_{ k_p}')
  \\
  & = &
  \frac{1}{2} \int_{Q_r\times Q_{r'}} du_1 \dots du_{k_p} du'_1 \dots du'_{k_p} 
  \, \big( \, \prod_{l=1}^{k_p} \, \delta(u_l - u'_l) \, \big) 
  \nonumber \\ 
  && \quad \quad \quad \quad \quad \quad
  (1 + u_1^2) \,  \sum_{j=1}^{k_p} \, \left( u_j^2 - (u'_j)^2 \right)  
   \widehat{\widetilde\gamma^{(k_p)}}(\uu_{ k_p};\uu_{ k_p}') 
  \label{eq-Ah1-rr-def-1}
\eeqn
for $Q_r,Q_{r'}\in \R^{dk_p}$ and $r,r'\in\Z^{dk_p}$. 

Moreover, for $Q_r,Q_{r'}\in \R^{d(2k_p-1)}$ and $r,r'\in\Z^{d(2k_p-1)}$, we let 
\eqn
  A_{b}^{(1)}(r,r') \, := \, 
%   \tr_1( \, K_1^{(1)} b_1^\pm \widetilde\gamma^{(2k_p)} \, + \, \mu \, K_1^{(2)} h_1^\pm \widetilde\gamma^{(k_p)} \, ) \,,
  \tr_1( \, K_1^{(1)} b_1^\pm \widetilde\gamma^{(2k_p-1)}_{r,r'} \, )
\eeqn
and 
\begin{align*} 
  A_{h}^{(2)}(r,r') \, &:= \,  \tr_1( \, K_1^{(2)} h_1^\pm \widetilde\gamma^{(k_p)}_{r,r'} \, ) \,,\\
                              \,  &= \,  \tr_1( \, K_1^{(2)} h_1^\pm \tr_{k_p+1\cdots2k_p-1}\widetilde\gamma^{(2k_p-1)}_{r,r'} \, ) \,,\\
\end{align*}
where we used admissibility to pass to the last line.
% and
% \eqn 
%   A_h^{(2)} \, := \, \tr_1( \, K_1^{(2)} h_1^\pm \widetilde\gamma^{(k_p)}\, ) \,,
% \eeqn
Also we let
\begin{align*} 
  A_b^{(2)}(r,r') & := \tr_1( \, K_1^{(2)} b_1^\pm \widetilde\gamma^{(2 k_p-1)}_{r,r'}\, )
  \\
  & = \frac{\mu}{p+2}\tr_1( \, B^+_{1;2,\dots,k_p} \, b_1^\pm \widetilde\gamma^{(2 k_p-1)}_{r,r'}\, )
  \\
  & = \frac{\mu}{p+2}
  \sum_{j=1}^{k_p} \tr_1( \, B^+_{1;2,\dots,k_p} \,  B^\pm_{j;k_p+1,\dots,2k_p -1}  
  \widetilde\gamma^{(2 k_p-1)}_{r,r'} \,) 
  \\
   & = \frac{\mu}{p+2}
  \sum_{j=1}^{k_p}
  \int dx_1 \dots dx_{k_p} dx'_1 \dots dx'_{k_p} \, 
  \delta(x_1 = \dots = x_{k_p} = x'_1 = \dots = x'_{k_p}) 
  \\ 
  & \quad \quad \quad \quad \quad \quad \quad \quad  \left( B^\pm_{j;k_p+1,\dots,2 k_p-1}  
  \widetilde\gamma^{(2 k_p-1)}_{r,r'} \right) (\ux_{k_p},\ux'_{k_p}) 
  \,,
\end{align*}
where we have adopted the notation
\eqn
	\lefteqn{
	\delta(x_1 = \dots = x_{k_p} = x'_1 = \dots = x'_{k_p}) \, 
	}
	\nonumber\\
	&&:= \, 
  \delta(x_1 - x'_1) \, 
  \prod_{\ell=2}^{k_p} \left( \, \delta(x_1-x_{\ell}) \, \delta(x_1 - x'_{\ell}) \, \right)
  \quad
\eeqn 
from \cite{chpatz1}.
\\

\noindent
\underline{\em Proof of $A_h^{(1)}(r,r')=0$.}
We recall that $Q_r\in\R^{dk_p}$ and $r,r'\in\Z^{dk_p}$. We have
\eqn  
  A_{h,+}^{(1)}(r,r') & := & 
  \frac{1}{2} \int_{Q_r\times Q_{r'}} du_1 \dots du_{k_p} du'_1 \dots du'_{k_p} 
  \, \big( \, \prod_{l=1}^{k_p} \, \delta(u_l - u'_l) \, \big) %\label{Ah+^1-1}
  \nonumber\\ 
  && \quad \quad \quad \quad \quad \quad
  (1 + u_1^2) \,  \sum_{j=1}^{k_p} \, u_j^2  \,
   \widehat{\widetilde\gamma^{(k_p)}}(\uu_{ k_p};\uu_{ k_p}') \label{Ah+^1-2}
  \\ 
  &=& \delta_{r,r'}\frac{1}{2} \int_{Q_r } du_1 \dots du_{k_p}   
  (1 + u_1^2) \,  \big( \, \sum_{j=1}^{k_p} \, u_j^2  \,\big)
   \widehat{\widetilde\gamma^{(k_p)}}(\uu_{ k_p};\uu_{ k_p})\,
  \nonumber\\
  &\leq& \sup_{\uu_{k_p}\in Q_r}\Big( \,\frac12 \, (1 + u_1^2) \,  \big( \, \sum_{j=1}^{k_p} \, u_j^2  \,\big) \, \Big)  \, \tr(\widetilde\gamma^{(k_p)})
  \nonumber\\
  &\leq& C(Q_r)
\eeqn
is finite, for every choice of $Q_r$. Therefore, 
\eqn 
  A_{h}^{(1)}(r,r') \, = \,  A_{h,+}^{(1)}(r,r') \, - \, A_{h,-}^{(1)}(r,r')  
  \, = \,   0 \, ,
\eeqn 
as claimed, since $A_{h,-}^{(1)}(r,r')= A_{h,+}^{(1)}(r,r') $ from direct comparison of terms in 
\eqref{Ah+^1-2} and \eqref{eq-Ah1-rr-def-1}.
  
\noindent
\underline{\em Proof of $A_b^{(2)}(r,r')=0$.}
We recall that in this situation, we have $Q_r\in\R^{d(2k_p-1)}$ and $r,r'\in\Z^{d(2k_p-1)}$.
Using the definition of $B^\pm_{j;k_p+1,\dots,2 k_p-1}$, we find that the term in $A_b^{(2)}(r,r')$
involving $B^+_{j;k_p+1,\dots,2 k_p-1}$ is given by
\eqn 
   \lefteqn{
  A_{b,+}^{(2)}(r,r') 
  }
  \nonumber\\
  & =& \frac{\mu}{p+2}
  \sum_{j=1}^{k_p}
  \int dx_1 \dots dx_{k_p} dx'_1 \dots dx'_{k_p} \, 
  \delta(x_1 = \dots = x_{k_p} = x'_1 = \dots = x'_{k_p}) 
  \nonumber\\ 
   && \quad \quad % [ \, 
  \widetilde\gamma^{(2k_p-1)}_{r,r'}(x_1,\dots, x_{k_p},\underbrace{x_j, \dots, x_j}_{k_p-1};
                     x'_1,\dots, x'_{k_p},\underbrace{x_j, \dots, x_j}_{k_p-1})
%   \nonumber\\
%   && \quad \quad \quad \quad 
%    - \widetilde\gamma^{(2k_p)}_{r,r'}(x_1,\dots, x_{k_p},\underbrace{x'_j, \dots, x'_j}_{k_p};
%                      x'_1,\dots, x'_{k_p},\underbrace{x'_j, \dots, x'_j}_{k_p}) \, ]
  \nonumber\\
  & =& \frac{\mu \, k_p}{p+2} 
  \int_{Q_r\times Q_{r'}} du_1 \dots du_{2k_p-1} du'_1 \dots du'_{2k_p-1} \, 
  \delta(\sum_{j=1}^{2k_p-1}(u_j-u_j')) 
  \nonumber\\ 
  && \quad \quad\quad \quad \quad \quad % [ \, 
  \widehat{\widetilde\gamma^{(2k_p-1)}}(\uu_{2k_p-1};\uu_{2k_p-1}')  \,.
%   \widehat{\widetilde\gamma^{(2k_p)}}(\uu_{2k_p};\uu_{2k_p}')  \, ] \,,
  \label{eq-cancel-finite-aux-3}
\eeqn
We need to verify that it is well-defined.

% Accordingly, let $\R^{2d k_p} \, = \, \bigcup_{r\in\Z^{2dk_p}} Q_r$ similarly as above, where
% $Q_r$ are disjoint cubes obtained from translating the 
% unit cube $[0,1)^{2dk_p}$ by  $r\in\Z^{2dk_p}$. 
% We want to prove that
% \eqn 
%   A_{b,+}^{(2)}(r,r') & :=&   \frac{\mu \, k_p}{p+2} 
%   \int_{Q_r\times Q_{r'}} du_1 \dots du_{2k_p} du'_1 \dots du'_{2k_p} \, 
%   \delta(\sum_{j=1}^{2k_p}(u_j-u_j')) 
%   \nonumber\\ 
%   && \quad \quad\quad \quad \quad \quad  
%   \widehat{\widetilde\gamma^{(2k_p)}}(\uu_{2k_p};\uu_{2k_p}')  
% \eeqn
% is finite, for every $(r,r')$.

To this end,  we  write
\eqn
	 \widehat{\widetilde\gamma^{(2k_p-1)}} (s;\uu_{2k_p-1} ; \uu_{2k_p-1}' ) \, = \, 
	\sum_j \lambda_j(s) \phi_j(\uu_{2k_p-1} ) \, \overline{\phi_j(\uu_{2k_p-1} ')} \,,
\eeqn
where $\{\phi_j\}$ is an orthonormal basis of $L^2(\R^{d(2k_p-1)})$, and 
$\sum_j\lambda_j(s)=1$, with $0\leq\lambda_j(s)\leq1$.
We observe that
\eqn
	\lefteqn{	
	\Big| \, \int_{Q_r\times Q_{r'}} d\uu_{2k_p-1}d\uu_{2k_p-1}' 
  \delta\big( \, \sum_{i=1}^{2k_p-1}(u_i-u_i') \, \big)   
  \phi_j(\uu_{2k_p-1} ) \, \overline{\phi_j(\uu_{2k_p-1} ')}  \, \Big| 
  }
  \nonumber\\
  &\leq&\frac12
  \int_{Q_r\times Q_{r'}} d\uu_{2k_p-1}d\uu_{2k_p-1}' 
  \delta\big( \, \sum_{i=1}^{2k_p-1}(u_i-u_i') \, \big)   
  \Big( \, |\phi_j(\uu_{2k_p-1} )|^2 \, + \, |\phi_j(\uu_{2k_p-1} ')|^2 \, \Big)
  \nonumber\\
  &=:&\frac12 \, \big( \, (i) \, + \, (ii) \, \big) \,,
\eeqn
where
\eqn
	(i) &=& \int_{Q_r\times Q_{r'}} d\uu_{2k_p-1}d\uu_{2k_p-1}' 
  \delta\big( \, \sum_{i=1}^{2k_p-1}(u_i-u_i') \, \big) \, |\phi_j(\uu_{2k_p} )|^2
  \nonumber\\
  &=&\int_{Q_r}d\uu_{2k_p-1}|\phi_j(\uu_{2k_p-1} )|^2 \, A(\sum_{i=1}^{2k_p-1}u_{i} ) 
  \nonumber\\
  &\leq & %\|\phi_j\|_{L^2}^2 
  \tr( \, \widetilde\gamma_{r,r}^{(2k_p-1)} \, )\, \sup_{\zeta\in\R^d}A(\zeta )   \,,
\eeqn
with 
\eqn
	A(\zeta) \, := \, \int_{ Q_{r'}}  d\uu_{2k_p-1}' \,
	\delta\big( \, \zeta-\sum_{i=1}^{2k_p-1}u_i' \, \big) \,.
\eeqn
We observe that this is the measure of the intersection  $Q_{r'}\cap L_\zeta$ where
\eqn
	L_\zeta \, := \, \{ \, \uu_{2k_p-1}'  \in \R^{d(2k_p-1)} \, | \, \sum_{i=1}^{2k_p-1}u_i' = \zeta \, \}
\eeqn
is an affine subspace of dimension $d(2k_p-2)$,
for $\zeta\in \R^d$. Clearly,
\eqn
	{\rm meas}( \, Q_{r'}\cap L_\zeta \, ) \,<\, {\rm meas}( \, B_{R}\cap L_\zeta \, )
	\, < \, C_0
\eeqn 
uniformly in $\zeta$, where we may take $B_R$ to be the smallest closed
ball (with $R$ denoting its radius) concentric with $Q_r$ such that $B_R\supset Q_r$. 
Indeed, the maximum of ${\rm meas}( \, B_{R}\cap L_\zeta \, )$ is attained when
$\zeta$ is such that the center of
$B_R$ is contained in $L_\zeta$, and corresponds to the volume of the 
$d(2k_p-2)$-dimensional ball of radius $R$.
We conclude that  
\eqn
	\sum_j \lambda_j(s) \|\phi_j\|_{L^2}^2 \, \sup_{\zeta}A(\zeta) 
	& \leq & C_0  \, \sum_j \lambda_j(s) 
	\nonumber\\
	&= & C_0 \,,
\eeqn
since $\|\phi_j\|_{L^2}^2=1$ for all $j$.
The same bound holds for the term $(ii)$.  This implies that 
\eqn 
  |\, A_{b,+}^{(2)}(r,r') \,| \, < \, C \, ( \, \tr( \, \widetilde\gamma_{r,r}^{(2k_p-1)} \, ) \, + \, 
  \tr( \, \widetilde\gamma_{r',r'}^{(2k_p-1)}) \, )
  \, < \,C' \,.
\eeqn
Moreover, one can straightforwardly see that including the term in $A_{b}^{(2)}(r,r') $ involving 
the operator $B^-_{j;k_p+1,\dots,2 k_p-1}$, one has $A_{b}^{(2)}(r,r')  = A_{b,+}^{(2)}(r,r') - A_{b,-}^{(2)}(r,r')$ where
$A_{b,-}^{(2)}(r,r')   =  A_{b,+}^{(2)}(r,r') $. Therefore,
\eqn 
  A_{b}^{(2)}(r,r') \, = \,    A_{b,+}^{(2)}(r,r') \, - \, A_{b,-}^{(2)}(r,r') \,   \, = \, 0 
\eeqn
is indeed satisfied.
\\

\noindent
\underline{\em Proof of $A_{h}^{(2)}(r,r')+\mu A_{b}^{(1)}(r,r')=0$.}
Here, we again have $Q_r\in\R^{d(2k_p-1)}$ and $r,r'\in\Z^{d(2k_p-1)}$.
We claim that  
\eqn\label{eq-Ah-cancel-1} 
  A_{h}^{(2)}(r,r')+\mu A_{b}^{(1)}(r,r')=0
\eeqn
holds. 

% Then,
% \eqn 
%   A_{h,b}^{(1,2)} \, = \, \sum_{r,r'}A_{h,b}^{(1,2)}(r,r')
% \eeqn
% where
% \eqn 
% %   & = & \tr_1( \, K_1^{(1)} b_1^\pm \widetilde\gamma^{(2k_p)} \, + \, \mu \, K_1^{(2)} h_1^\pm \widetilde\gamma^{(k_p)} \, ) 
% %   \nonumber\\
%   A_{h,b}^{(1,2)}(r,r')\, := \, \sum_{r,r'}
%   \tr_1( \, K_1^{(1)} b_1^\pm \widetilde\gamma^{(2k_p)}_{r,r'} \, + \, \mu \, K_1^{(2)} h_1^\pm \tr_{k_p+1\cdots2k_p}\widetilde\gamma^{(2k_p)}_{r,r'} \, )  \,.
% %   A_h(1;m) \, + \,\mu \,  A_b(1;m) \, = \, A_h^{(2)} \, + \, \mu \, A_b^{(1)} \, = \, 0 \, 
% \eeqn
% We claim that for every fixed $(r,r')$, the term in the sum is zero.

To this end, we note that
\begin{align}
  A_b^{(1)}(r,r') & = \, \frac{1}{2} \, \tr_1( \,(1 - \Delta_{x_1}) \, \tr_{2,\dots, k_p}
  \sum_{j=1}^{k_p} B^\pm_{j;k_p+1,\dots,2 k_p-1} \widetilde \gamma^{(2 k_p-1)}_{r,r'} \, )
  \nonumber\\
  & = \, \frac{1}{2} \, \sum_{j=2}^{k_p}\tr_{1,2,\dots, k_p}( \, (1 - \Delta_{x_1}) \,  
   B^\pm_{j;k_p+1,\dots,2 k_p-1} \widetilde \gamma^{(2 k_p-1)}_{r,r'} \, )
  \\
  &\quad\quad\quad
  \, + \, \frac{1}{2} \, \tr_{1,2,\dots, k_p}( \, (1- \Delta_{x_1}) \,  
   B^\pm_{1;k_p+1,\dots,2 k_p-1} \widetilde \gamma^{(2 k_p-1)}_{r,r'} \, )
  \nonumber\\
  & = 
  \, \frac{1}{2} \, \tr_{1,2,\dots, k_p}( \, (1 - \Delta_{x_1}) \,  
   B^\pm_{1;k_p+1,\dots,2 k_p-1} \widetilde \gamma^{(2 k_p-1)}_{r,r'} \, )\,
  \nonumber \\
  & = 
  \, \frac{1}{2} \, \int dx_1 \dots dx_{k_p} dx'_1 \dots dx'_{k_p} \, 
  \delta(x_1 - x_1') \cdots \delta(  x_{k_p} - x'_{k_p}) 
  \\ 
  & \quad \quad  \left( (1 - \Delta_{x_1}) B^\pm_{1;k_p+1,\dots,2 k_p-1}  
  \widetilde\gamma^{(2 k_p-1)}_{r,r'} \right) (\ux_{k_p},\ux'_{k_p}) \,.
\end{align}
Using the definition of $B^\pm_{1;k_p+1,\dots,2 k_p-1}$ and the identity (4.12) from \cite{chpatz1}, this equals
\begin{align}
	A_b^{(1)}(r,r')
  & = 
  \, \frac{1}{2} \, \int dx_1 \dots dx_{k_p} dx'_1 \dots dx'_{k_p} \, 
  \delta(x_1 - x_1') \cdots \delta(  x_{k_p} - x'_{k_p}) 
  \nonumber \\ 
  & \quad \quad  \left( \, 1 + \nabla_{x_1} \, \nabla_{x'_1} \, \right)
  \nonumber \\ 
  & \quad \quad \quad [ \, 
  \widetilde\gamma^{(2k_p-1)}_{r,r'}(x_1,\dots, x_{k_p},\underbrace{x_1, \dots, x_1}_{k_p-1};
                     x'_1,\dots, x'_{k_p},\underbrace{x_1, \dots, x_1}_{k_p-1})
  \nonumber \\
  & \quad \quad \quad \quad 
   - \widetilde\gamma^{(2k_p-1)}_{r,r'}(x_1,\dots, x_{k_p},\underbrace{x'_1, \dots, x'_1}_{k_p-1};
                     x'_1,\dots, x'_{k_p},\underbrace{x'_1, \dots, x'_1}_{k_p-1}) \, ]
  \nonumber \\
  & = 
  \, \frac{1}{2} \, \int dx_1 \dots dx_{k_p} dx'_1 \dots dx'_{k_p} \, 
  \delta(x_1 - x_1') \cdots \delta(  x_{k_p} - x'_{k_p})  
  \nonumber \\ 
  & \quad \quad [ \, \Delta_{x'_1} \, 
  \widetilde\gamma^{(2k_p-1)}_{r,r'}(x_1,\dots, x_{k_p},\underbrace{x_1, \dots, x_1}_{k_p-1};
                     x'_1,\dots, x'_{k_p},\underbrace{x_1, \dots, x_1}_{k_p-1})
  \nonumber \\
  & \quad \quad \quad \quad 
   - \Delta_{x_1} \widetilde\gamma^{(2k_p-1)}_{r,r'}(x_1,\dots, x_{k_p},\underbrace{x'_1, \dots, x'_1}_{k_p-1};
                     x'_1,\dots, x'_{k_p},\underbrace{x'_1, \dots, x'_1}_{k_p-1}) \, ].
  \label{comp-Ab1}
\end{align}
On the other hand, we consider
\begin{align} 
  A_h^{(2)}(r,r') & = - \, \frac{\mu}{p+2}
  \tr_1( \, B^+_{1;2,\dots, k_p}\sum_{j=1}^{k_p} 
  (\Delta_{x_j}-\Delta_{x_j'}) \tr_{k_p+1\cdots2k_p-1}\widetilde\gamma^{(2k_p-1)}_{r,r'} \, )
  \nonumber \\ 
  & = - \, \frac{\mu}{p+2}
   \, \int dx_1 \dots dx_{k_p} dx'_1 \dots dx'_{k_p} \, 
  \delta(x_1 = ... = x_{k_p} = x'_1 = ... = x'_{k_p}) 
  \nonumber \\ 
  & \quad \quad \sum_{j=1}^{k_p} ( \Delta_{x_j} - \Delta_{x'_j}) \,  
  \tr_{k_p+1\cdots2k_p-1}\widetilde{\gamma}^{(2k_p-1)}_{r,r'} \, (\ux_{2k_p-1}; \ux'_{2k_p-1}) 
\end{align}
By symmetry of $\widetilde{\gamma}^{(k_p)}$ with respect to the 
components of $\ux_{k_p}$ and $\ux'_{k_p}$, this yields
\begin{align}
	A_h^{(2)}(r,r')
  & =  - \, \frac{\mu \, k_p}{p+2} 
   \, \int dx_1 \dots dx_{k_p} dx'_1 \dots dx'_{k_p} \, 
  \delta(x_1 = ... = x_{k_p} = x'_1 = ... = x'_{k_p}) 
  \nonumber \\ 
  & \quad \quad ( \Delta_{x_1} - \Delta_{x'_1}) \,  
  \tr_{k_p\cdots2k_p-1}\widetilde{\gamma}^{(2k_p-1)}_{r,r'} \, (\ux_{2k_p-1};\ux'_{2k_p-1}) 
  \label{Ah2-sum}\\
  & =  - \, \frac{\mu}{2} 
  \, \int dx_1 \dots dx_{k_p} dx'_1 \dots dx'_{k_p} \, 
  \delta(x_1 - x_1') \cdots \delta(  x_{k_p} - x'_{k_p})
  \nonumber \\ 
  & \quad \quad [\,  
     \Delta_{x'_1} \widetilde{\gamma}^{(2k_p)}_{r,r'}(x_1,\dots, x_{k_p},\underbrace{x_1, \dots, x_1}_{k_p-1};
                     x'_1,\dots, x'_{k_p},\underbrace{x_1, \dots, x_1}_{k_p-1}) 
 \nonumber \\                    
  & \quad \quad \quad \quad
  -\Delta_{x_1} \widetilde{\gamma}^{(2k_p)}_{r,r'}(x_1,\dots, x_{k_p},\underbrace{x'_1, \dots, x'_1}_{k_p-1};
                     x'_1,\dots, x'_{k_p},\underbrace{x'_1, \dots, x'_1}_{k_p-1}) \,],
\label{comp-Ah2}
\end{align}
where we exchanged the roles of $(x_j,x_j')$ for $j=1,\dots,k_p$ with those of $(x_j,x_j')$ for $j=k_p+1,\dots,2k_p$.
% we used the admissibility of 
% $\Gamma = (\, \gamma^{(j)} \, )_{j \in \N}$ in order to obtain \eqref{comp-Ah2}. 

Similarly as above, one can prove that each term in the difference is separately finite, that is,
\eqn 
  A_{h}^{(2)}(r,r') \, = \,   A_{h,+}^{(2)}(r,r') \, - \, A_{h,-}^{(2)}(r,r')  
\eeqn
where both
\eqn 
  \lefteqn{
  A_{h,+}^{(2)}(r,r') 
  }
  \nonumber\\
  &:=&\frac{\mu \, k_p}{p+2}  \int_{Q_r\times Q_{r'}} du_1 \dots du_{2k_p-1} du'_1 \dots du'_{2k_p-1} \, 
  \delta(u_2 - u_2') \cdots \delta(  u_{k_p} - u'_{k_p}) \, u_1^2
  \nonumber \\ 
  &&\quad \quad \quad \quad \delta(u_1-u_1'+\sum_{j=k_p+1}^{2k_p-1}(u_j-u_j') ) \,
  \widehat{\widetilde{\gamma}^{(2k_p-1)}}(\uu_{2k_p-1};\uu_{2k_p-1}') 
\eeqn
and 
\eqn 
  \lefteqn{
  A_{h,-}^{(2)}(r,r') 
  }
  \nonumber\\
  &:=&\frac{\mu \, k_p}{p+2}  \int_{Q_r\times Q_{r'}} du_1 \dots du_{2k_p-1} du'_1 \dots du'_{2k_p-1} \, 
  \delta(u_2 - u_2') \cdots \delta(  u_{k_p} - u'_{k_p}) \, (u_1')^2
  \nonumber \\ 
  &&\quad \quad \quad \quad \delta(u_1-u_1'+\sum_{j=k_p+1}^{2k_p-1}(u_j-u_j') ) \,
  \widehat{\widetilde{\gamma}^{(2k_p-1)}}(\uu_{2k_p-1};\uu_{2k_p-1}') 
\eeqn
are bounded for any $(r,r')$. The proof is straightforwardly obtained from adapting the proof of $A_b^{(2)}(r,r')=0$,
by substituting $\widehat{\widetilde{\gamma}^{(2k_p-1)}}\rightarrow u_1^2\widehat{\widetilde{\gamma}^{(2k_p-1)}}$,
respectively  $\widehat{\widetilde{\gamma}^{(2k_p-1)}}\rightarrow (u_1')^2\widehat{\widetilde{\gamma}^{(2k_p-1)}}$
in the arguments given there.
The result is that
\eqn 
  A_{h,\pm}^{(2)}(r,r') \, < \, C \, \big( \, \tr( \, \widetilde\gamma_{r,r}^{(2k_p-1)} \, )
  \, + \, \tr( \, \widetilde\gamma_{r',r'}^{(2k_p-1)}) \, \big)
  \, < \,C' \,.
\eeqn
We shall not repeat the details here.

As a consequence, we obtain that
\eqn 
   \mu A_b^{(1)}(r,r') \, + \, A_{h}^{(2)}(r,r') \, = \, 0 \,,
\eeqn
by comparing terms in 
\eqref{comp-Ab1} and \eqref{comp-Ah2}.
This proves (\ref{eq-Ah-cancel-1}).
\\

%\noindent
%\underline{\em Conservation of higher order energies.}
Collecting the above results, we arrive at
\eqn
  \lefteqn{ 
  A (1;m)
  }
  \nonumber\\
  &=&\sum_{r,r'\in\Z^{dk_p}}A_h^{(1)}(r,r')  +  \sum_{r,r'\in\Z^{d(2k_p-1)}}\Big(\, \mu A_b^{(1)}(r,r') 
   +  A_h^{(2)}(r,r')  +  A_b^{(2)}(r,r')\, \Big) 
  \nonumber\\
  &=&0 \,,
\eeqn
and similarly, $A (\ell;m)=0$ for all $\ell=2,\dots,m$ (see \eqref{eq-Aellm-def-1}). Therefore,
\eqn   
  \partial_t\bra \cK^{(m)} \ket_{\Gamma(t)} 
  \, = \,
  \sum_{\ell=1}^m \Big[ \,A (\ell;m)
  %\, + \, 
  %\mu \, A_b(\ell;m) 
  \, \Big]  \, = \, 0 \,.
\eeqn 
Consequently, we have proved (\ref{eq-cKcons-id-1}). 
%\noindent
%\underline{\em A priori bound.}
%The proof of  \eqref{eq-Kxi-Gamma-aprioribd-0} follows from Theorem \ref{thm-globsp-encons-1}
%below, for  $\mu>0$ and
%$p\leq\frac{4}{d-2}$  (defocusing $p$-GP hierarchy). The argument is   the same also for
%$\mu<0$ and $p<p_{L^2}=\frac4d$ (focusing $L^2$ subcritical GP hierarchy).
\endprf

\section{Generalized Sobolev and Gagliardo-Nirenberg inequalities}
\label{sec-Sobolev-1}

As a preparation for Section \ref{sec-apriori}
where we use higher order energy conservation to obtain a
priori bounds on the $\frH_\xi^1$-norm of solutions of $p$-GP hierarchies, we present 
a generalization of Sobolev and Gagliardo-Nirenberg inequalities on
the level of density matrices.
We remark that W. Beckner recently derived a family of sharp estimates
in \cite{beck}  related to results of this type.

\begin{theorem}\label{eq-Sobmultidim-1}
(Sobolev inequality)
Assume that $f\in  H^\alpha(\R^{qd})$
for $\alpha> \alpha_0:=\frac{(q-1)d}{2q}<1$.
Then, there exists a constant $C(d,q,\alpha)$ such that 
\eqn
	\lefteqn{
  \Big(\int dx |f(\underbrace{x,\dots,x}_{q})|^2\Big)^{\frac12}
  }
  \nonumber\\
  & \leq & C(d,q,\alpha)
  \Big(\int dx_1 \, \cdots \, dx_q \, 
	\left| \, \bra\nabla_{x_1}\ket^{\alpha} \cdots \bra\nabla_{x_q}\ket^{\alpha} \,
	f(x_1,\dots,x_q)\right|^2\Big)^{\frac12}
	\nonumber\\
	&=&\| \, f \, \|_{ H^\alpha_{x_1,\dots,x_q}} 
\eeqn
for  $x_i\in\R^d$. In particular, $C(d,q,\alpha)<C'(d,q)|\alpha-\alpha_0|^{-q/2}$.
\end{theorem}

\prf
We use a Littlewood-Paley decomposition $1=\sum_j P_j$ where $P_j$ acts in
frequency space as multiplication with the characteristic function on the dyadic 
annulus $A_i:=\{\xi\in\R^d|2^j\leq|\xi|<2^{j+1}\}$, and $P_0$ is the characteristic
function on the unit ball.

Let $j_1,\dots,j_q\in\N_0$, and
\eqn
	\fjkl(x_1,\dots,x_q) \, := \, (P_{j_1}^{(1)}\cdots  P_{j_q}^{(q)}f)(x_1,\dots,x_q)
\eeqn
where the superscript in $P_i^{(m)}$ signifies that it acts on the $m$-th variable.

Then, clearly, the Fourier transform satisfies
\eqn
	\widehat\fjkl(\xi_1,\dots,\xi_q) \, = \, h_{j_1}(\xi_1) \cdots h_{j_q}(\xi_q) 
	\widehat\fjkl(\xi_1,\dots,\xi_q)
\eeqn
where $h_i$ are Schwartz class functions with $P_i h_i=P_i$.

We note that
\eqn
	h_j^\vee(x) & = & \int d\xi \, h_j(\xi) \, e^{2\pi i \xi x}
	\nonumber\\
	& = & 2^{jd }\int d\xi \, h_1(\xi) \, e^{2\pi i \xi(2^j x)}
	\nonumber\\
	& = & 2^{jd} \, h_1^\vee(2^j x) \,.
\eeqn
Therefore, $h_j^\vee$ is a smooth delta function with amplitude $2^{jd}$, and
supported on a ball of radius $2^{-j}$. In particular,
\eqn
	\| \, h_j^\vee \, \|_{L^\infty_x} \, \leq \, c \, 2^{jd}
	\; \; , \; \;
	\| \, h_j^\vee \, \|_{L^1_x} \, = \, \| \, h_1^\vee \, \|_{L^1_x} \, \leq \, c' \,,
\eeqn 
for constants $c$, $c'$ independent of $j$.
Because $h_j$ is an even function for
all $j$, it follows that $h_j^\vee\in\R$.  

Accordingly, performing the inverse Fourier transform,
\eqn 
	\fjkl(x_1,\dots,x_q)
	& = & \int dy_1 \, \cdots \,  dy_q \, \fjkl(y_1,\dots,y_q)  
	\\
	& &\quad\quad\quad\quad
	\, h_{j_1}^\vee(y_1+x_1) \, \cdots \, h_{j_q}^\vee(y_q+x_q) \,.
	\nonumber
\eeqn
In particular,
\eqn\label{eq-Sob-main-1}
	\lefteqn{
	\int dx \, | \fjkl(x,\dots,x)|^2 
	}
	\nonumber\\
	& = &
	\int dy_1 \, \cdots \,  dy_q \, dy_1' \, \cdots \,  dy_q' \,
	\fjkl(y_1,\dots,y_q)\overline{\fjkl(y_1',\dots,y_q')}
	\nonumber\\
	&&\quad\quad
	\int dx \, h_{j_1}^\vee(y_1+x) \, \cdots \, h_{j_q}^\vee(y_q+x)
	\nonumber\\
	&&\quad\quad\quad\quad\quad\quad
	 h_{j_1}^\vee(y_1'+x) \, \cdots \, h_{j_q}^\vee(y_q'+x) \,.
\eeqn
Using Cauchy-Schwarz only on $\fjkl\overline{\fjkl}$, this is bounded by
\eqn\label{eq-Sob-main-2}
	\int dx \, | \fjkl(x,\dots,x)|^2 & \leq &
	\int  dy_1 \, \cdots \,  dy_q \,
	| \, \fjkl(y_1,\dots,y_q) \, |^2
	\nonumber\\
	&& 
	\int dx \, \int  dy_1' \, \cdots \,  dy_q' \,
	| \, h_{j_1}^\vee(y_1+x) \, \cdots \, h_{j_q}^\vee(y_q+x) 
	\nonumber\\
	&&\quad\quad\quad\quad\quad\quad
	h_{j_1}^\vee(y_1'+x) \, \cdots \, h_{j_q}^\vee(y_q'+x) \, | \,.
\eeqn 
Thus, integrating out $y_1',\dots,y_q'$ and using $\|h_j^\vee\|_{L^1_x}<c'$, 
\eqn
	\int dx \, | \fjkl(x,\dots,x)|^2 & \leq &C \,
	\int dy_1 \, \cdots \,  dy_q \,
	| \, \fjkl(y_1,\dots,y_q) \, |^2
	\nonumber\\
	&& 
	\int dx \, | \,
	h_{j_1}^\vee(y_1+x) \, \cdots \, h_{j_q}^\vee(y_q+x) \, |  \,.
\eeqn
Now, we assume without any loss of generality that $j_i\leq j_q$ for all $i< q$. Then,
\eqn
	\int dx \, 
	| \, h_{j_1}^\vee(y_1+x) \, \cdots \, h_{j_q}^\vee(y_q+x) \, |
	& \leq & 
	\|h_{j_1}^\vee\|_{L^\infty_x}\cdots\|h_{j_{q-1}}^\vee\|_{L^\infty_x}\|h_{j_q}^\vee\|_{L^1_x}
	\nonumber\\
	& \leq &
	 2^{(j_1+\cdots+j_{q-1})d} \,  c' \,.
\eeqn
Now, since by assumption, $j_i\leq j_q$ for all $i< q$, 
\eqn\label{eq-max-jq-1}
	j_1+\cdots+j_{q-1} \, \leq \, \frac{q-1}{q}(j_1+\cdots+j_{q}) \,.
\eeqn
Therefore,
\eqn
	\lefteqn{
	\int dx \, |\fjkl(x,\dots,x)|^2 
	}
	\nonumber\\
	& \leq &C \, 2^{2(j_1+\cdots+j_{q})\alpha_0}
	\int dy_1 \, \cdots \,  dy_q \,
	| \, \fjkl(y_1,\dots,y_q) \, |^2 \,,
	\label{eq-SobLq-aux-0}
\eeqn
where 
\eqn
	\alpha_0 \, := \, \frac{(q-1)d}{2q} \,.
\eeqn
We thus find that
\eqn\label{eq-SobLq-aux-1}
	\lefteqn{
	\|\,  f(x,\dots,x) \, \|_{L^2_x(\R^d)}
	}
	\nonumber\\
	&=&\|\, \sum_{j_1,\dots,j_q} \fjkl(x,\dots,x) \, \|_{L^2_x(\R^d)}
	\nonumber\\
	&\leq&\sum_{j_1,\dots,j_q}\Big(\int dx \, |f_{j_1,\dots,j_q}(x,\dots,x)|^2 \Big)^{\frac12}
	\nonumber\\
	& \leq &C \,\sum_{j_1,\dots,j_q} 2^{ (j_1+\cdots+j_{q})\alpha_0 }
	\Big( \int dy_1 \, \cdots \,  dy_q \,
	| \, f_{j_1,\dots,j_q}(y_1,\dots,y_q) \, |^2 \Big)^{\frac12}
	\nonumber\\
	& = &C \,\Big(\sum_{j_1,\dots,j_q} 2^{ -2(j_1+\cdots+j_{q})\e }\Big)^{\frac12}
	\Big(\sum_{j_1,\dots,j_q} 2^{ 2(j_1+\cdots+j_{q})\alpha } 
	\| \, f_{j_1,\dots,j_q} \, \|_{L^2_{x_1,\dots,x_q}}^2 \Big)^{\frac12}
	\nonumber\\
	&=&C_\e \, \| \, f \, \|_{ H^\alpha_{x_1,\dots,x_q}} \,.
	\label{almostSob}
\eeqn
for $\alpha=\alpha_0+\e$, and any $\e>0$. 
We note that from the above,  $C_\e\leq C\e^{-q}$ follows immediately.
This is the asserted result.
\endprf

We immediately obtain the following Gagliardo-Nirenberg type inequality.

\begin{theorem}\label{thm-GaglNir-1}
(Gagliardo-Nirenberg inequality)
Assume that $f\in  H^1(\R^{qd})$ 
and $\alpha>\alpha_0:=\frac{(q-1)d}{2q}<1$. Then, there exists a constant $C(d,q,\alpha)$ such that
\eqn\label{eq-GaglNir-1}
	\Big(\int dx \, | \, f(\underbrace{x,\dots,x}_{q}) \, |^2 \Big)^{\frac12}\, \leq \, 
	C(d,q,\alpha) \,  \| \, f \, \|_{ H^1_{x_1,\dots,x_q}}^{ \alpha} 
	\| \, f \, \|_{L^2_{x_1,\dots,x_q}}^{ 1-\alpha }
\eeqn
where $x_i\in\R^d$. In particular, $C(d,q,\alpha)<C'(d,q)|\alpha-\alpha_0|^{-q/2}$.
\end{theorem}

\prf
From \eqref{almostSob}, the H\"older estimate yields
\eqn
	\lefteqn{
	\|\,  f(x,\dots,x) \, \|_{L^2_x(\R^d)}
	}
   \nonumber\\
  &\leq&  \sum_{j_1,\dots,j_q} \|\, \fjkl(x,\dots,x) \, \|_{L^2_x(\R^d)} 
	\nonumber\\
	& \leq &C \,\sum_{j_1,\dots,j_q} 2^{ (j_1+\cdots+ j_q)\alpha_0 }
	\| \, f_{j_1,\dots,j_q}  \, \|_{L^2} 
	\nonumber\\
	& = & C \,\sum_{j_1,\dots,j_q} 2^{ (j_1+\cdots+ j_q)\alpha_0 } 
	\| \, f_{j_1,\dots,j_q} \, \|_{L^2}^{ \alpha} 
	\| \, f_{j_1,\dots,j_q} \, \|_{L^2}^{ 1-\alpha }
	\nonumber\\
	& \leq & C \, \Big[\sum_{j_1,\dots,j_q} \Big(2^{ (j_1+\cdots+ j_q)(\alpha_0+\delta) } 
	\| \, f_{j_1,\dots,j_q} \, \|_{L^2}^{ \alpha}\Big)^{\frac1\alpha} \Big]^{ \alpha }
	\nonumber\\
	&&\quad\quad\quad
	\Big[\sum_{j_1,\dots,j_q}  \Big( \, 2^{ -(j_1+\cdots+ j_q)\delta } 
	\| \, f_{j_1,\dots,j_q} \, \|_{L^2}^{1-\alpha}\Big)^{\frac{1}{1-\alpha}}\Big]^{ 1-\alpha } \,.
\eeqn
Letting   $\e=\delta>0$ and $1>\alpha\geq\alpha_0+2\e$, where $\alpha_0=\frac{(q-1)d}{2q}<1$,
this is bounded by
\eqn
	& \leq & C\, \Big[\Big(\sum_{j_1,\dots,j_q} 2^{ -2(j_1+\cdots+ j_q)\frac{\e}{\alpha} } 
	\Big)^{\frac1{2}} 
	\Big(\sum_{j_1,\dots,j_q} 2^{ 2(j_1+\cdots+ j_q)\frac{\alpha_0+\delta+\e}{\alpha} } 
	\| \, f_{j_1,\dots,j_q} \, \|_{L^2}^{ 2}\Big)^{\frac1{2}} \Big]^{ \alpha }
	\nonumber\\
	&&\quad\quad\quad
	\Big[\Big(\sum_{j_1,\dots,j_q} 2^{ -2(j_1+\cdots+ j_q)\frac{\delta}{1-\alpha} } 
	\Big)^{\frac1{2}} 
	\Big(\sum_{j_1,\dots,j_q}    \,
	\| \, f_{j_1,\dots,j_q} \, \|_{L^2}^{2}\Big)^{\frac{1}{2}}\Big]^{ 1-\alpha }
	\nonumber\\
	& \leq & C\, \Big[\Big( \frac{\e}{\alpha}   
	\Big)^{-\frac q{2}} 
	\Big(\sum_{j_1,\dots,j_q}  
	\| \, f_{j_1,\dots,j_q} \, \|_{H^{\frac{\alpha_0+\delta+\e}{\alpha}}}^{ 2}\Big)^{\frac1{2}} \Big]^{ \alpha }
	\nonumber\\
	&&\quad\quad\quad
	\Big[\Big( \frac{\delta}{1-\alpha} 
	\Big)^{-\frac q{2}} 
	\Big(\sum_{j_1,\dots,j_q}    \,
	\| \, f_{j_1,\dots,j_q} \, \|_{L^2}^{2}\Big)^{\frac{1}{2}}\Big]^{ 1-\alpha }
	\nonumber\\
	&\leq&C\Big(\alpha^\alpha(1-\alpha)^{1-\alpha}\frac1\e\Big)^{q/2} \, \| \, f \, \|_{ H^1_{x_1,\dots,x_q}}^{ \alpha} 
	\| \, f \, \|_{L^2_{x_1,\dots,x_q}}^{ 1-\alpha }
	\nonumber\\
	&\leq&C\e^{-q/2} \, \| \, f \, \|_{ H^1_{x_1,\dots,x_q}}^{ \alpha} 
	\| \, f \, \|_{L^2_{x_1,\dots,x_q}}^{ 1-\alpha }\,.
\eeqn
Here we observed that $\frac1\alpha,\frac1{1-\alpha}>1$ are H\"older conjugate exponents,
and we used that $\sup_{\tau\in[0,1]}\tau^\tau=1$.
\endprf

We note that $\alpha_0=\frac d4$ for the cubic, $p=2$, and $\alpha_0=\frac{3d}8$ for
the quintic case, $p=4$.
Accordingly,  one may choose $\e\geq\frac15$ in both cases (i.e., $\e$ is not very small),
where $d\leq 3$ for the cubic, and $d\leq 2$ for the quintic case.

\section{A priori energy bounds}
\label{sec-apriori}

In this section we use higher order energy functionals to obtain three types of bounds: 
\begin{enumerate} 
\item A priori energy bounds for both focusing and defocusing energy-subcritical p-GP hierarchies (Subsection \ref{sec-bd-foc-defoc}). 
\item A priori $H^1$ bounds for defocusing energy subcritical p-GP hierarchies (Subsection \ref{sec-globalwp-1}).
\item A priori $H^1$ bounds for $L^2$-subcritical focusing p-GP hierarchies (Subsection \ref{sec-globwp-L2subcrit-1}).
\end{enumerate}

First, we present a priori energy bounds which are valid for solutions of the focusing and defocusing energy-subcritical 
$p$-GP hierarchies. 

\subsection{A priori energy bounds for focusing and defocusing energy-subcritical p-GP hierarchies} 
\label{sec-bd-foc-defoc} 
\begin{theorem}
\label{thm-bd-foc-defoc}
Let $p < \frac{4}{d-2}$.
If $\Gamma(t)\in\frH^{1}_{\xi'}$
%for $t\in[0,T]$, 
is a solution to the $p$-GP hierarchy (focusing or defocusing), 
%with initial data  $\Gamma_0 \in \frH_{\xi'}^1$ for  
%\eqn
%	\xi \, \leq \, (1+\frac{2}{p+2}\CSob (d,p))^{-\frac{1}{k_p}} \, \xi' \,.
%\eeqn
then the a priori bound
\eqn
	\sum_{m\in\N}(2\xi)^m \bra \cK^{(m)} \ket_{\Gamma(t)} 
	\, \leq \,  \|\Gamma(t)\|_{\frH_{\xi'}^1}
\label{eq-Kxi-Gamma-aprioribd-0}
\eeqn
holds for all $\xi$ satisfying 
\eqn
	\xi \, \leq \, (1+\frac{2}{p+2}\CSob (d,p))^{-\frac{1}{k_p}} \, \xi' \,
\eeqn
and all $t\in\R$.
\end{theorem}

\prf 
Let $p < \frac{4}{d-2}$. We shall use the Sobolev inequalities for the GP hierarchy,
Theorem \ref{eq-Sobmultidim-1} to bound the interaction energy 
by the kinetic energy:
\eqn \label{gwp-corSob}
	\tr_1 ( \, B_{1;2,\dots,k_p} \widetilde\gamma^{(k_p)}\, )
	\, \leq \, \CSob (d,p) \, \tr_{1,\dots,k_p}( \, S^{(k_p,1)} \widetilde\gamma^{(k_p)} \, ) \,.
\eeqn
To see this, we write $\widetilde{\gamma}^{(k_p)}$ as
\eqn 
  \widetilde{\gamma}^{(k_p)}(\ux_{k_p},\ux_{k_p}') \, = \, \sum_j \lambda_j 
  \big|\phi_j(\ux_{k_p})\ket \bra \phi_j(\ux_{k_p}')\big| \,,
\eeqn	
with respect to an orthonormal basis $(\phi_j)_j$,
where $\lambda_j\geq0$ and $\sum_j\lambda_j=1$. Then we have 
\eqn \label{normcon-basis}  
  \tr_1( \, B_{1;2,\dots,k_p}^+ \widetilde{\gamma}^{(k_p)}\, ) 
  \, = \,  \sum_j \lambda_j 
  \int dx |\phi_j(\underbrace{x,\dots,x}_{k_p})|^2 \,.
\eeqn
Now Theorem  \ref{eq-Sobmultidim-1} (with $q=k_p$ and $\alpha = 1$) implies that 
for $p < \frac{4}{d-2}$ we have
\eqn
  \lefteqn{
  \tr_1( \, B_{1;2,\dots,k_p}^+ \widetilde{\gamma}^{(k_p)}\, ) 
  }
  \nonumber\\
  & \leq & C_{Sob} \, \sum_j 
  \lambda_j \|\phi_j\|_{H^1_{\ux_{k_p}}}^2
  \nonumber\\
  & \leq & C_{Sob} \, \sum_j 
  \lambda_j \|\phi_j\|_{\frh^1_{\ux_{k_p}}}^2
  \nonumber\\
 & = & C_{Sob} \, \tr_{1,\dots,k_p}( \, S^{(k_p,1)} \widetilde\gamma^{(k_p)} \,).
\eeqn
Accordingly, \eqref{gwp-corSob} implies 
\eqn
	\tr_1 ( \, K_1 \widetilde\gamma^{(k_p)}\, )
	\, \leq \, ( \, \frac12+\frac{1}{p+2} \CSob (d,p) \, ) \, 
	\tr_{1,\dots,k_p}( \, S^{(k_p,1)} \widetilde\gamma^{(k_p)} \, ) \,.
\eeqn
By iteration, we obtain that 
\eqn 
  \bra \cK^{(m)} \ket_{\Gamma }  
  & \leq & 
  ( \, \frac12+\frac{1}{p+2}\CSob(d,p) \, )^m \, \tr_{1,\dots,mk_p}( \, S^{(m k_p,1)} \gamma^{(m k_p)} \, ) \,.
\eeqn
Therefore, 
\eqn
	\sum_{\ell}(2\xi)^\ell \bra \cK^{(\ell)} \ket_{\Gamma} 
	& \leq &
	\sum_{\ell}\big( \, ( \, 1+\frac{2}{p+2}\CSob(d,p) \, )^{\frac{1}{k_p}} \, \xi \, \big)^\ell
	\| \, \gamma^{(\ell)} \, \|_{\frh_\ell^1}
	\nonumber\\
	& \leq & \| \, \Gamma \, \|_{\frH_{\xi'}^1}
\eeqn
for all $\xi$ satisfying 
\eqn 
  \xi \, \leq \, ( \, 1+\frac{2}{p+2}\CSob(d,p) \, )^{-\frac{1}{k_p}} \, \xi' \,.
\eeqn
Hence, the claim follows.
\endprf

\subsection{A priori $H^1$ bounds for defocusing energy subcritical GP hierarchies}
\label{sec-globalwp-1} 

For energy subcritical, defocusing GP hierarchies, we can now
deduce a priori energy bounds as follows.

\begin{theorem}
\label{thm-globsp-encons-1}
Assume that $\mu=+1$ (defocusing), 
$p < \frac{4}{d-2}$,  
and that $\Gamma(t)\in\frH^{1}_\xi$, $t\in[0,T]$, 
is a positive semidefinite solution of 
the $p$-GP hierarchy with initial data  $\Gamma_0 \in \frH_{\xi'}^1$  
for  
\eqn
	\xi \, \leq \, (1+\frac{2}{p+2}\CSob (d,p))^{-\frac{1}{k_p}} \, \xi' \,.
\eeqn
Then, one finds 
\eqn
	\|\Gamma(t)\|_{\frH_{\xi}^1}   & \leq &\sum_{m\in\N}(2\xi)^m \bra \cK^{(m)} \ket_{\Gamma(t)}  
	\nonumber\\
	& = &
	\sum_{m\in\N}(2\xi)^m \bra \cK^{(m)} \ket_{\Gamma_0} 
	\, \leq \,  \|\Gamma_0\|_{\frH_{\xi'}^1} \label{eq-bd-defoc}
\eeqn
for all $t\in[0,T]$.
\end{theorem}

\prf
We first note that for the defocusing $p$-GP hierarchy, 
it follows immediately from Theorem \ref{thm-Enconserv-1} that
\eqn\label{eq-H1apriori-bd-1}
  \|\Gamma(t)\|_{\frH_{\xi}^1}  & \leq &
  \sum_{m\in\N}(2\xi)^m \bra \cK^{(m)} \ket_{\Gamma(t)} 
  \nonumber\\
  &=&
  \sum_{m\in\N}(2\xi)^m \bra \cK^{(m)} \ket_{\Gamma_0} 
\eeqn
for $t\in[0,T]$.  
The first inequality is obtained by discarding all of the (positive) interaction energies
in $\bra \cK^{(m)} \ket_{\Gamma(t)}$.
 
 Subsequently, we use the a priori bound \eqref{eq-Kxi-Gamma-aprioribd-0}
 derived in Theorem 
\ref{thm-bd-foc-defoc} to obtain \eqref{eq-bd-defoc}.
\endprf

\subsection{A priori energy bounds for $L^2$ subcritical focusing GP hierarchies}
\label{sec-globwp-L2subcrit-1} 

In this subsection, we prove a priori $\frH_\xi^1$-bounds 
for focusing $L^2$ subcritical $p$-GP hierarchies,
$p<p_{L^2}=\frac{4}{d}$.
The defocusing case is already contained in Theorem \ref{thm-globsp-encons-1}. 
The analogous result for the NLS is well-known, and is based on the use of energy
conservation where the $H^1$ norm (if large) of the solution is seen to dominate over
the potential energy, via the Gagliardo-Nirenberg inequality. 
We obtain the following similar result in the context
of the $L^2$-subcritical $p$-GP hierarchy.

\begin{theorem}\label{thm-L2subcrit-1}
Let $p<p_{L^2}=\frac4d$ ($L^2$ subcritical).
Moreover, let $\alpha>\alpha_0:=\frac{(k_p-1)d}{2k_p}$ and $\alpha k_p <1$,
where $k_p=1+\frac p2$, and $\alpha<1$.  Let 
\eqn 
 	 D\, : = \, D(\alpha, p, d, |\mu|)  \, = \,   \left( 1 - |\mu|  \frac{C_0(\alpha)}{1-4^{-(1-\alpha k_p)} } \right),
\label{eq-defD-1} 
\eeqn
where $C_0(\alpha)$ is characterized in \eqref{eq-Bgamma-GaglNir-1}.

Assume that  $\Gamma(t)\in\frH_\xi^{1}$ 
is a positive semidefinite solution of the focusing ($\mu<0$) $p$-GP hierarchy 
for $t\in [0,T]$,
given initial data $\Gamma(0)=\Gamma_0\in\frH_{\xi'}^{1}$ where 
\eqn\label{eq-xixiprime-Sob-0-1}
	\xi \, \leq \,   \frac{1}{D} \,  \big( \, 1+\frac{2}{p+2}\CSob (d,p) \, \big)^{-\frac{1}{k_p}} \, \xi' \,.
\eeqn 
If $\mu<0$ is such that  
\eqn\label{eq-mu-uppbd-0-5}
  |\mu| \, < \,  \frac{ 1-4^{-(1-\alpha k_p)} }{ C_0(\alpha)  } \,,
\eeqn 
then the a priori bound
\eqn
  \| \, \Gamma(t) \, \|_{\frH_{\xi}^1}  
  \, & \leq & \, \sum_{m=1}^\infty \, (2 D \, \xi)^m \, \bra \, \cK^{(m)} \, \ket_{\Gamma(t)} 
  \label{eq-H1-aprioribd-firstline-1} \\
  \, & =  & \, \sum_{m=1}^\infty \, (2 D \, \xi)^m \, \bra \, \cK^{(m)} \, \ket_{\Gamma_0}
  \label{eq-5xiKm-apriori-0} \\
  &\leq& \|\Gamma_0\|_{\frH_{\xi'}^1}  
\label{eq-H1-aprioribd-1}
\eeqn
holds for all $t\in [0,T]$.  
\end{theorem}

\prf 
First, we observe that it follows immediately from \eqref{eq-Kxi-Gamma-aprioribd-0}
that \eqref{eq-5xiKm-apriori-0} implies \eqref{eq-H1-aprioribd-1}.

Next, we show that $\|\Gamma(t)\|_{\frH_\xi^1}$ is bounded by 
the right hand side in \eqref{eq-H1-aprioribd-firstline-1}. 
Given $\alpha>\alpha_0=\frac{(k_p-1)d}{2k_p}$ where $\alpha_0<1$, 
we infer from Theorem \ref{eq-Sobmultidim-1} that
\eqn\label{eq-Bgamma-GaglNir-1}
	\tr_1( \, B_{1;2,\dots,k_p}^+ \gamma^{(k_p)}\, )
	\, \leq \, C_0(\alpha) \, \tr( \,   S^{(k_p,\alpha)} \, \gamma^{(k_p)} \, )  \,,
\eeqn
as follows.
We write $\gamma^{(k_p)}$ 
with respect to an orthonormal eigenbasis $(\phi_j)_j$,
\eqn 
  \gamma^{(k_p)}(\ux_{k_p},\ux_{k_p}') \, = \, \sum_j \lambda_j 
  \big|\phi_j(\ux_{k_p})\ket \bra \phi_j(\ux_{k_p}')\big| \,,
\eeqn	
where $\lambda_j\geq0$ and $\sum_j\lambda_j=1$, so that
\eqn 
  \tr_1( \, B_{1;2,\dots,k_p}^+ \gamma^{(k_p)}\, ) 
  \, = \,  \sum_j \lambda_j 
  \int dx |\phi_j(\underbrace{x,\dots,x}_{k_p})|^2 \,.
\eeqn
Then Theorem \ref{eq-Sobmultidim-1} (with $q=k_p$)
implies that 
\eqn
  \lefteqn{
  \tr_1( \, B_{1;2,\dots,k_p}^+ \gamma^{(k_p)}\, ) 
  }
  \nonumber\\
  & \leq & C_0(\alpha) \, \sum_j \lambda_j  
  \|\phi_j\|_{\frh^\alpha_{\ux_{k_p}}}^2
  \nonumber\\
  & \leq &C_0(\alpha) \,  \tr( \,   S^{(k_p,\alpha)} \, \gamma^{(k_p)} \, ) \,,
\eeqn
which is what we claimed in \eqref{eq-Bgamma-GaglNir-1}.
%Hence from \eqref{eq-Bgamma-GaglNir-1}, we have that  
%\eqn  
%  \int dx \, \gamma^{(k_p)}(x,\dots,x;x,\dots,x) 
%  \, \leq \, C_0(\alpha) \,   \tr ( \, S^{(k_p,\alpha)} \, \gamma^{(k_p)} \, ) \,.
%  \label{eq-freq-rest-Bgamma-1}
%\eeqn 
%for $\alpha>\alpha_0=\frac{(p+1)d}{2p+4}$.

Next, we recall the definition of the operators 
\eqn  
  K_\ell \, = \, K_\ell^{(1)} \, + \, K_\ell^{(2)}
\eeqn
where
\eqn
  K_\ell^{(1)} \, := \,  \frac{1}{2} \, ( 1 \, - \, \Delta_{x_\ell}) \, \tr_{\ell+1,\dots,\ell+\frac p2}  
\eeqn
and
\eqn
  K_\ell^{(2)} \, := \, \frac{\mu}{p+2} \,  B_{\ell;\ell+1,\dots,\ell+\frac p2}^+ \,,
\eeqn
for $\ell\in\N$.
Moreover, we have  
\eqn
  \cK^{(m)} & := & K_{1} 
  K_{k_p + 1}  \cdots K_{(m-1) k_p + 1} \,,
\eeqn
and we proved that
\eqn
   \bra\cK^{(m)}\ket_{\Gamma(t)} \, := \,  \tr_{1,k_p + 1 ,2k_p + 1,\dots,(m-1)k_p + 1} 
   ( \, \cK^{(m)}\gamma^{(m k_p)}(t) \, ) 
\eeqn
is a conserved quantity, for every $m\in\N$, provided that $\Gamma(t)$ solves the
$p$-GP hierarchy. 

In order to simplify the presentation below, we
introduce the notation 
\eqn
	\tr^{1,m} \, : = \, \tr_{1,k_p + 1 ,2k_p + 1,\dots,(m-1)k_p + 1} \,.
\eeqn  
In the $L^2$ subcritical case, where $p<p_{L^2}=\frac4d$, we will now use the 
sequence of conserved quantities $(\bra\cK^{(m)}\ket_{\Gamma(t)} )_{m\in\N}$
to obtain an a priori bound on $\|\Gamma(t)\|_{\frH_\xi^1}$, for $\xi>0$ sufficiently
small.

We define the Fourier multiplication operator $P_{j;i}^{(r)}$ acting on $f:(\R^{d})^{ k_p}\rightarrow\R$ by
\eqn  
  (\widehat{P_{j;i}^{(r)}f})(\uxi_r)  
  \, := \, p_j(\xi_i) \,
  \widehat{f}(\uxi_r) \,,
\eeqn 
with symbols $p_j(\xi_i)$ that are smooth and supported
in ${2 \over 3} 2^j < |\xi_i| < 3 (2^j)$, for $j > 1$ and in 
$|\xi_1| \leq 3$ when $j=0$. 
Also we define
$P_{\leq j;i}^{(r)} = \sum_{k \leq j} P_{k;i}^{(r)}$ (and denote the corresponding symbol 
by $p_{\leq j; i}$).

\noindent
Moreover, we define  the Fourier multiplication operator $P_{j;i}^{(r),\leq}$ via 
 \eqn  
  (\widehat{P_{j;i}^{(r), \leq}f})(\uxi_r)  
  \, := \, p_j(\xi_i) 
  \Big[\prod_{\ell=2\atop l\neq i}^r p_{\leq j}(\xi_\ell)  \Big] \,
  \widehat{f}(\uxi_r) \,,
\eeqn 
and the Fourier multiplication operator $P_{j;i i'}^{(r)}$ acting on $\gamma^{(r)}$ by
$$
P_{j;i i'}^{(r)} = P_{j;i}^{(r), \leq} \; P_{j;i'}^{(r),\leq}.
$$
%Also we note that
%\eqn 
%   \tr ( \, P_{j;ii'}^{(k_p)} \,   \gamma^{(k_p)} \, ) \, = \, \delta_{i i'} \tr ( \, P_{j;ii}^{(k_p)} \,   \gamma^{(k_p)} \, ) \,.
%\eeqn 

Then we have
%\eqn 
%  \sum_{i=1}^{k_p} \tr ( \, P_{j;ii}^{(k_p)} \,   \gamma^{(k_p)} \, )
  %& = &  \sum_{i=1}^{k_p} \int d\uxi_{k_p} \, P_{j;i}^{(r)}(\uxi_{k_p})  \, \widehat{\gamma^{(k_p)}}(\uxi_{k_p};\uxi_{k_p})
  %\nonumber\\
%  & = & k_p \, \tr ( \, P_{j;11}^{(r)} \,   \gamma^{(k_p)} \, )
%  \nonumber\\
% & = & k_p \, \int d\uxi_{k_p} \, P_j(\xi_1) \, P_{\leq j}(\xi_2) \,\cdots\, P_{\leq j}(\xi_{k_p}) \, 
%  \widehat{\gamma^{(k_p)}}(\uxi_{k_p};\uxi_{k_p})
%  \nonumber\\
% & \leq & k_p \, \int d\xi_1 \, P_j(\xi_1) \, \int d\xi_2\cdots d\xi_{k_p} \, 
%  \widehat{\gamma^{(k_p)}}(\uxi_{k_p};\uxi_{k_p})
%  \nonumber\\
%  &=& k_p \, \int d\xi_1 \, P_j(\xi_1) \, 
% \widehat{\gamma^{(1)}}(\xi_1;\xi_1)
%  \nonumber\\
% & = & k_p \, \tr ( \, P_{j}  \,   \gamma^{(1)} \, ) \,,
%\eeqn 
%using symmetry on the first line, and admissibility to pass to the second last line.
%Therefore,
\eqn 
   \lefteqn{
   \tr ( \, S^{(k_p,\alpha)} \, \gamma^{(k_p)} \, )
  }
  \nonumber\\
  & \leq & \sum_{j=0}^{\infty}\sum_{i=1}^{k_p} \tr ( \, P_{j;ii}^{(k_p)} \, S^{(k_p,\alpha)} \, \gamma^{(k_p)} \, )
  \nonumber\\
& = & \sum_{j=0}^{\infty} k_p \tr ( \, P_{j;ii}^{(k_p)} \, S^{(k_p,\alpha)} \, \gamma^{(k_p)} \, )
  \label{eq-foc-lo-adm1}\\
& = & \sum_{j=0}^{\infty} k_p 
\int d\uxi_{k_p} p_j(\xi_1) \left( \prod_{\ell = 2}^{k_p} p_{\leq j} (\xi_{\ell}) \right) 
\left( \prod_{k = 1}^{k_p}  (1+ |\xi_k|^2)^{\alpha} \right) \widehat{\gamma}(\uxi_{k_p}; \uxi_{k_p})
\nonumber\\
& \leq & \sum_{j=0}^{\infty} k_p  \, (1+ 2^{2j})^{\alpha k_p} \, 
\int d\xi_1 \, p_j(\xi_1) \;  \int d\xi_2...d\xi_{k_p} \,
\widehat{\gamma}(\uxi_{k_p}; \uxi_{k_p})
\nonumber\\
& = & \sum_{j=0}^{\infty} k_p  \, (1+ 2^{2j})^{\alpha k_p} \, 
\int d\xi_1 p_j(\xi_1) \widehat{\gamma}(\xi_1; \xi_1)
\label{eq-foc-lo-adm2}\\
& = & \sum_{j=0}^{\infty} k_p \, (1 + 2^{2j})^{\alpha k_p} \tr ( \, P_{j;1}  \,   \gamma^{(1)} \, ) \,, 
\label{eq-freq-rest-Bgamma-10}
\eeqn  
where to obtain \eqref{eq-foc-lo-adm1} we used symmetry and to obtain \eqref{eq-foc-lo-adm2} we used admissibility of 
$\gamma^{(k_p)}$. 

Hence, we find
\eqn 
  \lefteqn{
  2 \, \tr( \, K_1\gamma^{(k_p)} \, )
  }
  \nonumber\\
  &\geq& 2 \, \tr( \, K_1^{(1)} \, \gamma^{(k_p)} \, ) 
  \, - \, \frac{2 |\mu| C_0(\alpha)}{p+2} \, \tr( \, S^{(k_p,\alpha)} \, \gamma^{(k_p)} \, )
  \label{eq-foc-lo-comb-sob}\\
  & \geq &  \tr( \,  S^{(1,1)} \gamma^{(1)} \, ) 
  \, - \, \frac{2  |\mu| C_0(\alpha) k_p}{p+2} \, \sum_{j=0}^{\infty} (1 + 2^{2j})^{\alpha k_p} \, \tr( \, P_{j;1}^{(1)} \,   \gamma^{(1)} \, )
  \label{eq-foc-lo-comb-usingpriv}\\ 
  &\geq&  \tr( \,  S^{(1,1)} \gamma^{(1)} \, ) 
  - \, |\mu| C_0(\alpha)  \, \sum_{j=0}^{\infty} (1 + 2^{2j})^{-(1-\alpha k_p)} \, \tr( (1+ 2^{2j}) \, P_{j;1}^{(1)} \,   \gamma^{(1)} \, )
  \nonumber \\
  &\geq&  \tr( \,  S^{(1,1)} \gamma^{(1)} \, ) 
  - \, \left( \sup_{j \geq 0} \tr( (1+ 2^{2j}) \, P_{j;1}^{(1)} \,   \gamma^{(1)} \, ) \, \right)
  \, |\mu| C_0(\alpha) \, \sum_{j=0}^{\infty} (1 + 2^{2j})^{-(1-\alpha k_p)}
  \nonumber \\
  & \geq & \left( 1 - |\mu| C_0(\alpha) \sum_{j=0}^{\infty} 2^{-2j(1-\alpha k_p)}  \right) \, \tr( \,  S^{(1,1)} \gamma^{(1)} \, ) 
  \label{eq-foc-lo-cheapLP} \\
  & = & D \; \tr( \,  S^{(1,1)} \gamma^{(1)} \, ) \,, 
  \label{eq-foc-lowbd-1}
\eeqn 
with 
$$
 	D\, :=  \, D(\alpha, p, d, |\mu|) \, = \,   
 	\left( 1 - |\mu|  \frac{C_0(\alpha)}{1 - 4^{-(1-\alpha k_p)} } \right) \, .
$$ 
Here to obtain \eqref{eq-foc-lo-comb-sob} we used  \eqref{eq-Bgamma-GaglNir-1}, 
to obtain \eqref{eq-foc-lo-comb-usingpriv} we used  \eqref{eq-freq-rest-Bgamma-10}, 
to obtain \eqref{eq-foc-lo-cheapLP} we used the inequality about Littlewood-Paley operators
$$ 
	\sup_{j \geq 0} \tr( (1+ 2^{2j}) \, P_{j;1}^{(1)} \,   \gamma^{(1)} \, ) 
	\, \leq \, \tr( \,  S^{(1,1)} \gamma^{(1)} \, ),$$
and to obtain \eqref{eq-foc-lowbd-1} we used the fact that we consider the $L^2$ subcritical problem, hence
 $\alpha k_p < 1$. 

%At this point, we observe that on the $L^2$-critical level, we can always pick $\alpha>\alpha_0$ in such a way that 
%$\alpha k_p<1$. Hence, there exists an explicit constant $J=J(\alpha,d,p)\in\N$ such that
%\eqn
%  1 \, -
%  \, \frac{4  |\mu| C_0(\alpha) k_p}{p+2} \,  
%(1 + 2^{2j})^{-(1 - \alpha k_p)} \, \geq \, \frac12
%\eeqn 
%for all $j\geq J$, so that in particular,
%\eqn\label{eq-Jexp-def-1}
%1 +   2^{2J} \, = \,\Big( \, \frac{8  |\mu| C_0(\alpha) k_p}{p+2} \, \Big)^{\frac{1}{1-\alpha k_p}} \,.
%\eeqn 
%This implies
%\eqn 
% \tr( \, K_1\gamma^{(k_p)} \, ) 
%  &\geq&\frac14\sum_{j=J}^\infty (1 + 2^{2j}) \,  \tr ( \, P_{j;1}^{(1)}\gamma^{(1)} \, ) \, 
%  \nonumber\\
% &\geq&\frac1{5 } \, \tr ( \,P_{\geq J;1} S^{(1,1)}\gamma^{(1)} \,) \,.
%  \label{eq-K1gamma-lowbd-1}
%\eeqn
%for all $j\geq J=J(\alpha,d,p)$, using admissibility on the last line. 

Accordingly, we conclude that if $|\mu|$ is such that $D(\alpha, p, d, |\mu|) > 0$,
i.e.
\eqn 
|\mu| \, < \, \frac{ 1-4^{-(1-\alpha k_p)} }{ C_0(\alpha)  } 
\label{eq-foc-mu} 
\eeqn 
then  
\eqn 
  \|\gamma^{(1)}\|_{\frh^1}
  & = &  \tr ( \, S^{(1,1)} \, \gamma^{(1)} \,) 
  \nonumber\\
  &\leq& \frac{2}{D} \;  \tr( \, K_1\gamma^{(k_p)} \, ) \,.
\eeqn 
Next, we generalize this inequality to the higher order energy functionals.

To this end, we observe that by iterating  \eqref{eq-foc-lowbd-1} we obtain 
\eqn \label{eq-cKm-apriori-aux-1}
  \lefteqn{
  \bra\cK^{(m)}\ket_{\Gamma(t)} \, = \, \tr^{1,m}( \, K_{1} 
  K_{k_p + 1}  \cdots K_{(m-1) k_p + 1} \gamma^{(m k_p)}\, )
  }
  \nonumber\\
  &\geq&\frac1{D} \, \tr^{1,m}( \, K_{1} 
  K_{k_p + 1}  \cdots K_{(m-2) k_p + 1} K_{(m-1) k_p + 1}^{(1)}\gamma^{(m k_p)}\, ) 
  \nonumber\\
  &\geq&\cdots \cdots \, 
  \nonumber\\
  &\geq &  \Big(\frac1{D} \Big)^m \, \tr^{1,m} \Big(\, K_{1}^{(1)} 
  K_{k_p + 1}^{(1)}  \cdots  K_{(m-1) k_p + 1}^{(1)}\gamma^{(m k_p)}\, \Big) 
  \nonumber\\
  &=&
  \Big(\frac1{2D} \Big)^m\tr^{1,m}( \, S^{(m k_p,1)}\gamma^{(m k_p)}\, ) 
  \nonumber\\
  &=&\Big(\frac1{2D} \Big)^m \, \| \, \gamma^{(m)}\, \|_{\frh^1}\,,
  \quad\quad
\eeqn
using the admissibility of $\gamma^{(m k_p)}$ in order to obtain the last line.
%where 
%\eqn 
% (P_{>J}^{(m)}\gamma^{(m)})(\uxi_m;\uxi_m) \, := \,   \Big[\prod_{\ell=1}^m P_{> J}(\xi_\ell)\Big]
% \, \gamma^{(m)}(\uxi_m;\uxi_m) \,.
%\eeqn 
%$$ 
%P_{\geq J}^{(m)} = \prod_{\ell = 1}^m  P_{\geq J, \ell}.
%$$

%Let the constant $K_{\Gamma_0}$ satisfy
%\eqn 
% \bra\cK^{(m)}\ket_{\Gamma(t)} \, = \, \bra\cK^{(m)}\ket_{\Gamma(0)}
%  \,  \leq \, K_{\Gamma_0}^m \,
%\eeqn 
%for all $m$. 
%Then,

%\eqn 
%  \| \, \gamma^{(m)}(t) \, \|_{\frh^1} & \leq &
%  ( \, 2D \, K_{\Gamma_0} \, )^m \, ,
%  \nonumber\\
%  &\leq&\Big( \, \big( \, \frac{8  |\mu|  C_0(\alpha) k_p}{p+2} \, \big)^{\frac{1}{1-\alpha k_p}} 
% \, + \, 5 \, K_{\Gamma_0} \, \Big)^m\,,
%\eeqn 
%where the right hand side  
%only depends on $\alpha>\alpha_0$, and $d$, $p$, as well as on the initial data $\Gamma_0$.
Consequently, 
%recalling that $k_p=\frac{p+2}{2}$,
\eqn 
  \| \, \Gamma(t) \, \|_{\frH_{\xi}^1} 
    \, & \leq & \, \sum_{m=1}^\infty \, \xi^m \,  \| \gamma^{(m)} \|_{{\frh}^1}
  \nonumber \\
  \, & \leq & \, \sum_{m=1}^\infty \, (2D\, \xi)^m \,  \bra\cK^{(m)}\ket_{\Gamma(t)} 
  \nonumber \\
%  \,&  =  & \, \sum_{m=1}^\infty \, \xi^m \, 
% \big( \, 2D  \, K_{\Gamma_0} \, \big)^m  
  \, & = & \, \sum_{m=1}^\infty \, (2D\, \xi)^m \,  \bra\cK^{(m)}\ket_{\Gamma(0)} \,.
\eeqn
This concludes the proof. 
%and \eqref{eq-pr-foc-bd-concl} is finite for any $\xi>0$ satisfying 
%\eqn\label{eq-L2subcrit-xi-bd-1}
%  \xi \, < \, \big( \, 2D \, K_{\Gamma_0} \, \big)^{-1} \,.
%\eeqn 
%In particular, the upper bound in \eqref{eq-L2subcrit-xi-bd-1} is independent of the  
%local wellposedness time $T$, and the inequality can accordingly be extended to all $t\in\R$,
%as will be addressed below.
%
%Next, we address the case $\mu<0$ with 
%\eqn 
%  |\mu| \, < \, \frac{p+2}{8   C_0(\alpha) k_p} \, = \, \frac1{4 C_0(\alpha)} \,.
%\eeqn 
%In this situation, \eqref{eq-Jexp-def-1} is easily seen to be replaced by
%\eqn\label{eq-Jexp-def-2}
%  2^{2J} \, \leq \,\big( \,  4 |\mu| C_0(\alpha) k_p \, \big)^{\frac{1}{1-\alpha k_p}} \, < \, 1 \,.
%\eeqn 
%Accordingly, $J=0$, so that \eqref{eq-cKm-apriori-aux-1} implies
%\eqn \label{eq-cKm-apriori-aux-2}
%  \lefteqn{
% \bra\cK^{(m)}\ket_{\Gamma(t)} \, = \, \tr^{1,m}( \, K_{1} 
%  K_{k_p + 1}  \cdots K_{(m-1) k_p + 1} \gamma^{(m k_p)}\, )
%  }
% \nonumber\\
%  &\geq& \Big(\frac1{5} \Big)^m \, \| \, P_{>0}^{(m)}\gamma^{(m k_p)}\, \|_{\frh^1}
%  \nonumber\\
%  &=& \Big(\frac1{5} \Big)^m \, \| \,  \gamma^{(m)}\, \|_{\frh^1}\,,
%  \quad\quad
%\eeqn
%so that instead of \eqref{eq-gammam-L2subcrit-uppbd-1},
%we have
%\eqn\label{eq-gammam-L2subcrit-uppbd-2} 
%   \| \, \Gamma(t) \, \|_{\frH_\xi^1} 
%  \, \leq \, \sum_{m\geq1} (5\xi)^m \, \bra\cK^{(m)}\ket_{\Gamma(t)}   \,,
%\eeqn 
%as claimed.
\endprf

\section{Global well-posedness of solutions in $\cH^1_\xi$}
\label{sec-globalwp-1-1} 

In this section, we will use the
the higher order energy functionals in order to enhance local to global wellposedness for 
solutions in the spaces $\cH_\xi^1$ constructed in \cite{chpa2}, for initial data in $\frH_\xi^1$.  
The local well-posedness result proven in \cite{chpa2} has the following form.

\begin{theorem}
\label{thm-localwp-TTstar-1-1}
Let
\eqn\label{eq-alphaset-def-0-1} 
	\alpha \, \in \, \alphaset(d,p)
	\, := \,  \left\{
	\begin{array}{cc}
	(\frac12,\infty) & {\rm if} \; d=1 \\ 
	(\frac d2-\frac{1}{2(p-1)}, \infty) & {\rm if} \; d\geq2 \; {\rm and} \; (d,p)\neq(3,2)\\
	\big[1,\infty) & {\rm if} \; (d,p)=(3,2) \,.
	\end{array}
	\right.
\eeqn
Then, there exists a constant $0<\eta<1$ such that for $0<\xi\leq\eta\xi'\leq1$, 
there exists a constant $T_0(d,p,\xi,\xi')>0$ such that
the following holds.  Let $I:=[0,T]$ for $0<T<T_0(d,p,\xi,\xi')$.  
Then, there exists a unique solution
$\Gamma\in L^\infty_{t\in I}\cH_{\xi}^\alpha$ of the $p$-GP hierarchy, with 
\eqn 
  \| \, \opB\Gamma \, \|_{ L^1_{t\in I}\cH_{\xi}^\alpha } \, < \, 
  C(T,\xi,\xi',d,p) \, \|\Gamma_0\|_{\cH_{\xi'}^\alpha} \,,
\eeqn
in the space 
\eqn\label{eq-Wspace-def-1-1}
	{\mathcal W}^\alpha(I,\xi) \, = \, \{ \, \Gamma\in L^\infty_{t\in I}\cH_{\xi}^\alpha \, | \,
	\opB^{+}\Gamma \, , \, \opB^{-}\Gamma \in  L^2_{t\in I}\cH_{\xi}^\alpha \, \}
\eeqn
for the initial condition $\Gamma(0)=\Gamma_0\in\cH_{\xi'}^\alpha$. 
\end{theorem}

% 
% The key improvement of this local well-posedness result
% over the one established in \cite{chpa2} consists of the fact that the 
% initial condition and the solution are in the same space $\cH_{\xi}^\alpha$.
% In \cite{chpa2}, the initial data $\Gamma_0$ was required to belong to $\cH_{\xi_1}^\alpha$
% for some $\xi_1>0$, while the solution $\Gamma(t)$ was shown to belong to
% $\cH_{\xi_2}^\alpha$, for some $0<\xi_2<\xi_1$. For the proof of Theorem \ref{thm-localwp-TTstar-1},
% we refer to \cite{chpa3}.
% \\

We will next prove the following theorem.

\begin{theorem}
\label{thm-globsp-encons-1-1}
Assume that one of the two following cases is given:
\begin{itemize}
 \item  Energy subcritical, defocusing $p$-GP hierarchy with $p<\frac{4}{d-2}$ and  $\mu=+1$. $\xi$, $\xi'$ satisfy 
    \eqref{eq-xixiprime-Sob-0}.
 \item  $L^2$ subcritical, focusing $p$-GP hierarchy with  $p<\frac4d$ and $\mu<0$ with 
  $|\mu| \, < \,  \frac{ 1-4^{-(1-\alpha k_p)} }{ C_0(\alpha)  }$
   where $C_0(\alpha)$ is characterized in  \eqref{eq-Bgamma-GaglNir-1}.
   $\xi$, $\xi'$ satisfy \eqref{eq-xixiprime-Sob-0-5}.
\end{itemize}
Then, there exists $T>0$ such that for $I_j:=[jT,(j+1)T]$, with $j\in\Z$, there exists  
a unique global solution $\Gamma\in \cup_{j\in\Z}{\mathcal W}^1(I_j,\xi)$ of the $p$-GP hierarchy
with initial condition $\Gamma(0)=\Gamma_0\in\cH_{\xi'}^1$, satisfying
\eqn
	\|\Gamma(t)\|_{\cH_{\xi}^1} \, \leq \, C  \|\Gamma_0\|_{\frH_{\xi'}^1} 
\eeqn
for all $t\in\R$, if $\Gamma(t)$ is positive semidefinite for all $t\in I_j$, $j\in\Z$.
\end{theorem}

\prf
We need to prove that the higher order energy functionals are conserved for solutions 
in Theorem \ref{thm-localwp-TTstar-1-1}

For $m\in\N$ fixed, we consider $\gamma^{(mk_p)}$. Letting $(\phi_j)$ be an orthonormal basis of $L^2(\R^{dmk_p})$, we
define
\eqn 
  \gamma^{(mk_p)}_J(t) \, := \,\sum_{j\in J}\lambda_j(t)\phi_j(t,\ux_{mk_p})
  \overline{\phi_j(t,\ux_{mk_p}')}
\eeqn
for any finite subset $J\subset\Z$, with $0\leq\lambda_j(t)\leq1$. 

Furthermore, we decompose  
\eqn
	\R^{jmd} \, = \, \bigcup_{r\in\Z^{jmd}} Q_r
\eeqn 
into disjoint cubes $Q_r$ obtained from translating the
unit cube $[0,1)^{jmd}$ by  $r\in\Z^{jmd}$. 
We then define
\eqn 
  \widetilde\gamma^{(j)}_{r,r';J} \, := \, P_{Q_r,Q_{r'}}(\widetilde\gamma^{(j)}_J)
\eeqn
where $P_{Q_r,Q_{r'}}$ is the Fourier multiplication operator with symbol given by the characteristic
function of $Q_r\times Q_{r'}$. 

Accordingly, letting $\ell=1$, we let
\eqn
  \widetilde\gamma^{(k_p)}_{r,r';J}(\ux_{ k_p};\ux_{ k_p}')
  \, := \, \tr_{k_p+1,\dots,mk_p}( \, K_{k_p+1}\cdots K_{(m-1)k_p+1} \gamma^{(mk_p)}_{r,r';J} \, )
\eeqn
and
\eqn 
  \lefteqn{
  \widetilde\gamma^{(2k_p-1)}_{r,r';J}(\ux_{ k_p},\uy_{ k_p-1};\ux_{ k_p}',\uy_{ k_p-1}')
  }
  \\
  & := & \tr_{k_p+1,\dots,mk_p}( \, K_{k_p+1}\cdots K_{(m-1)k_p+1})\} \gamma^{(mk_p + \frac{p}{2})}_{r,r';J} \, )
  (\ux_{ k_p},\uy_{ k_p-1};\ux_{ k_p}',\uy_{ k_p-1}')
  \nonumber
\eeqn
where $y_i=x_{m k_p+i}$ and $y_i'=x_{m k_p+i}'$, for $i\in\{1,\dots,k_p-1\}$. 

Then, we define, similarly to the notation used in the proof of Theorem \ref{thm-Enconserv-1},
\eqn\label{eq-Asigmai-def-cH1-1}
  A_h^{(1)}(r,r';J) & := & \tr_1( \, K_1^{(1)} h_1^\pm \widetilde \gamma^{(k_p)}_{r,r';J} \, ) 
  \nonumber\\ 
  A_{b}^{(1)}(r,r';J) & := &   
  \tr_1( \, K_1^{(1)} b_1^\pm \widetilde\gamma^{(2k_p-1)}_{r,r';J} \, )
  \nonumber\\
  A_{h}^{(2)}(r,r';J) & := &  \tr_1( \, K_1^{(2)} h_1^\pm \tr_{k_p+1\cdots2k_p-1}\widetilde\gamma^{(2k_p-1)}_{r,r';J} \, ) 
  \nonumber\\
  A_b^{(2)}(r,r';J) & :=& \tr_1( \, K_1^{(2)} b_1^\pm \widetilde\gamma^{(2 k_p-1)}_{r,r';J}\, ) 
\eeqn

we find that each of these terms can be written in the form
\eqn 
  A_{\sigma}^{(i)}(r,r';J) \, = \, A_{\sigma,+}^{(i)}(r,r';J) \, - \, A_{\sigma,-}^{(i)}(r,r';J)
\eeqn
for $\sigma\in\{h,b\}$ and $i\in\{1,2\}$. 
Following the arguments leading to the proof of Theorem \ref{thm-Enconserv-1}, each term in the difference can be estimated by
\eqn 
  \lefteqn{
  | \, A_{\sigma,\nu}^{(i)}(r,r';J) \, | 
  }
  \nonumber\\
  & < & C \, \big( \, \tr( \, \widetilde\gamma_{r,r;J}^{(j)} \, )
  \, + \, \tr( \, \widetilde\gamma_{r',r';J}^{(j)}) \, \big) 
  \nonumber\\
  & < &C \,|J|^{1/2} \, \big( \,  [\, \tr( \,| \widetilde\gamma_{r,r;J}^{(j)}|^2 \, )\, ]^{1/2}
  \, + \,  \tr( \, [\widetilde\gamma_{r',r';J}^{(j)}) \, ]^{1/2} \, \big) 
  \nonumber\\
  & < &2 \, C \, |J|^{1/2} \, \| \, \widetilde\gamma^{(j)} \, \|_{L^2(\R^{dj}\times \R^{dj})}
   \nonumber\\
  & < &2 \, C \, |J|^{1/2} \, \| \, \gamma^{(mk_p)} \, \|_{H^1(\R^{dmk_p}\times \R^{dmk_p})} \,,
\eeqn
where $\nu\in\{+,-\}$ and $j \in \{k_p, 2k_p -1\}$.
By the local wellposedness of the solution in $\cH_\xi^1$, the last line is bounded. 
Therefore, all terms in \eqref{eq-Asigmai-def-cH1-1} are well-defined, and similarly as in
the proof of Theorem \ref{thm-Enconserv-1}, they cancel, 
\eqn 
  A_{h}^{(1)}(r,r';J) \, + \, A_{h}^{(2)}(r,r';J) \, + \, \mu A_{b}^{(1)}(r,r';J) \, + \, A_{b}^{(2)}(r,r';J) \, = \, 0
\eeqn
for all $r,r'\in\Z^{dmk_p}$, and all $J$ with $|J|<\infty$. This implies that as in \eqref{eq-partialt-cKm-def-1},
\eqn\label{eq-partialt-cKm-def-1-1}
  \lefteqn{
  \partial_t\bra \cK^{(m)} \ket_{\Gamma(t)} 
  }
  \nonumber\\ 
  &=&\sum_{J\in\cI}\sum_{r,r'}\big( \, A_{h}^{(1)}(r,r';J)  +  A_{h}^{(2)}(r,r';J) 
   +  \mu A_{b}^{(1)}(r,r';J)  +  A_{b}^{(2)}(r,r';J) \, \big) 
  \nonumber\\
  & = & 0
\eeqn
where $\Z=\cup_{J\in\cI}J$ is an arbitrary decomposition of $\Z$ into mutually disjoint discrete intervals $J$ of finite size.

Accordingly, if the solution $\Gamma(t)\in\cH_\xi^1$ obtained from Theorem \ref{thm-localwp-TTstar-1-1} has 
initial data in the subspace $\Gamma(0)=\Gamma_0\in\frH_{\xi'}^1\subset \cH_{\xi'}^1$, then
\eqn 
  \bra \cK^{(m)} \ket_{\Gamma(t)} & = & \bra \cK^{(m)} \ket_{\Gamma_0}  \, + \, \int_0^t ds \, \partial_s \bra \cK^{(m)} \ket_{\Gamma(s)} 
  \nonumber\\
  &=& \bra \cK^{(m)} \ket_{\Gamma_0} \,,
\eeqn
for all $m\in\N$. That is, the higher order energy functionals are conserved.

In the case $p<\frac{4}{d-2}$ (energy subcritical), and $\mu=+1$ (defocusing), 
we choose $\xi_1,\xi$ such that
\eqn
	0 \, < \, \xi_1 \, \leq \, \eta \, \xi \, \leq \, 
	\eta \, (1+\frac{2}{p+2}\CSob (d,p))^{-\frac{1}{k_p}} \, \xi' \,.
\eeqn
where $0<\eta<1$ is the constant in Theorem \ref{thm-localwp-TTstar-1-1}.
Next, we pick 
\eqn 
  T \, := \, \min\{ \,T(d,p,\xi_1,\xi) \, , \, T(d,p,\xi,\xi') \, \} \,,
\eeqn
where $T(d,p,\xi,\xi')$ is defined as in Theorem \ref{thm-localwp-TTstar-1-1}.  

Then, Theorem \ref{thm-globsp-encons-1} implies that the a priori bound
\eqn
	\|\Gamma(t)\|_{\cH_{\xi}^1}&\leq&\|\Gamma(t)\|_{\frH_{\xi}^1} 
	\, \leq \, \sum_{m\in\N}(2\xi)^m \bra \cK^{(m)} \ket_{\Gamma(t)}  
	\nonumber\\
	& = &
	\sum_{m\in\N}(2\xi)^m \bra \cK^{(m)} \ket_{\Gamma_0} 
	\, \leq \,  \|\Gamma(0)\|_{\frH_{\xi'}^1}
\eeqn
holds for $t\in I_0=[0,T]$. 

Next, Theorem \ref{thm-localwp-TTstar-1-1} implies that there exists a unique solution in 
${\mathcal W}^1(I_1,\xi_1)$ for $t\in I_1=[T,2T]$, 
for the initial condition $\Gamma(T)\in\cH_{\xi}^1$.
Therefore, we have for every $m\in\N$,
\eqn 
   \bra \cK^{(m)} \ket_{\Gamma(t)} \, = \, \bra \cK^{(m)} \ket_{\Gamma(T)}  \, = \, \bra \cK^{(m)} \ket_{\Gamma_0} 
\eeqn
for all $t\in I_1$. Since these are $t$-independent $\C$-numbers, this implies that 
\eqn
	\|\Gamma(t)\|_{\cH_{\xi}^1}&\leq&\|\Gamma(t)\|_{\frH_{\xi}^1}  
	\, \leq \, \sum_{m\in\N}(2\xi)^m \bra \cK^{(m)} \ket_{\Gamma(t)}  
	\nonumber\\
	& = &
	\sum_{m\in\N}(2\xi)^m \bra \cK^{(m)} \ket_{\Gamma_0} 
	\, \leq \,  \|\Gamma(0)\|_{\frH_{\xi'}^1}
\eeqn
for $t\in I_1$. In particular, this is true for the same value of $\xi$ as above, for the case of the interval $I_0$.  
Thus, by bootstrapping, we find that instead of $\Gamma(t)\in\cH_{\xi_1}^1$, one
in fact has $\Gamma(t)\in\cH_{\xi}^1$ for all $t\in[T,2T]$, where $\xi_1\leq \eta\xi$.

Again invoking Theorem \ref{thm-localwp-TTstar-1-1} and repeating the above, we furthermore conclude that 
there exists a unique solution  
$\Gamma\in{\mathcal W}^1(I_1,\xi_1)$ for $t\in I_2=[2T,3T]$ for the initial condition $\Gamma(2T)\in\cH_{\xi}^1$.
Accordingly, higher order energy
conservation shows that  $\Gamma(t)\in\cH_{\xi}^1$ for all $t\in I_2$.

Iterating this argument, we have proved that there exists  
a unique global solution $\Gamma\in \cup_{j\in\Z}{\mathcal W}^1(I_j,\xi_1)$ of the $p$-GP hierarchy
with initial condition $\Gamma(0)=\Gamma_0\in\cH_{\xi'}^1$, satisfying
\eqn
	\|\Gamma(t)\|_{\cH_{\xi_1}^1} \, < \, \|\Gamma(t)\|_{\cH_{\xi}^1} \, \leq \, \|\Gamma_0\|_{\frH_{\xi'}^1} 
\eeqn
for all $t\in\R$.

In the case $p<\frac4d$ ($L^2$ subcritical), and $\mu<0$ (focusing) with 
\eqn 
 |\mu| \, < \,  \frac{ 1-4^{-(1-\alpha k_p)} }{ C_0(\alpha)  }
\eeqn  
we use that
\eqn\label{eq-H1-aprioribd-1-1}
   \| \, \Gamma(t) \, \|_{\cH_{\xi}^1}  & \leq &
  \| \, \Gamma(t) \, \|_{\frH_{\xi}^1}  \, \leq \,   
  \sum_{m\in\N}(2D\,\xi)^m \bra \cK^{(m)} \ket_{\Gamma(t)}
  \nonumber\\
  &=&
  \sum_{m\in\N}(2D\,\xi)^m \bra \cK^{(m)} \ket_{\Gamma(0)}
  \, \leq \,  \|\Gamma(0)\|_{\frH_{\xi'}^1}
\eeqn
for all $t\in[0,T]$, and for  $\xi$, $\xi'$ satisfying \eqref{eq-xixiprime-Sob-0-5}.  
Accordingly, we can repeat the arguments for the case $p<\frac{4}{d-2}$ and $\mu=+1$.  
This completes the proof.
\endprf

\subsection*{Acknowledgements}
We are grateful to I. Rodnianski and M. Visan for very useful comments.
We thank W. Beckner, B. Erdogan and N. Tzirakis for helpful discussions. 
The work of T.C. was supported by NSF grant DMS 0704031 / DMS-0940145, DMS-1009448,
and DMS-1151414 (CAREER).
The work of N.P. was supported NSF grant number DMS 0758247, DMS 1101192 
and an Alfred P. Sloan Research Fellowship.

\end{document}